\documentclass[11pt]{article}

\usepackage[utf8]{inputenc}
\usepackage{multirow}
\usepackage{booktabs}
\usepackage{todonotes}
\usepackage{float}
\usepackage[numbers]{natbib}

\usepackage{amsmath}
\usepackage{amssymb}
\usepackage{amsfonts}
\usepackage{bbm}
\usepackage{graphicx}
\setkeys{Gin}{draft=false}

\usepackage{subcaption}
\usepackage{url}

\usepackage{mathsymbols}

\DeclareMathOperator{\E}{\mathbb{E}}
\DeclareMathOperator{\I}{\mathbb{I}}

\begin{document}
\title{Beyond Personalization: Research Directions in Multistakeholder Recommendation}

\author{Himan Abdollahpouri$^1$ \and Gediminas Adomavicius$^2$ \and Robin Burke$^1$ \and 
Ido Guy$^3$ \and Dietmar Jannach$^4$ \and
Toshihiro Kamishima$^5$ \and Jan Krasnodebski$^6$ \and Luiz Pizzato$^7$}
\date{$^1$University of Colorado, Boulder, $^2$University of Minnesota, 
$^3$eBay, Inc., 
$^4$University of Klagenfurt,
$^5$National Institute of Advanced Industrial Science and Technology, Japan,
$^6$Expedia Group,
$^7$Accenture
}


\maketitle

\begin{abstract}
    Recommender systems are personalized information access applications; they are ubiquitous in today's online environment, and effective at finding items that meet user needs and tastes. As the reach of recommender systems has extended, it has become apparent that the single-minded focus on the user common to academic research has obscured other important aspects of recommendation outcomes. Properties such as fairness, balance, profitability, and reciprocity are not captured by typical metrics for recommender system evaluation. The concept of multistakeholder recommendation has emerged as a unifying framework for describing and understanding recommendation settings where the end user is not the sole focus. This article describes the origins of multistakeholder recommendation, and the landscape of system designs. It provides illustrative examples of current research, as well as outlining open questions and research directions for the field.

\end{abstract}

\section{Introduction}
\label{sec:introduction}
\noindent  Recommender systems provide personalized information access, supporting e-commerce, social media, news, and other applications where the volume of content would otherwise be overwhelming. They have become indispensable features of the Internet age, found in systems of many kinds. Even search engines, the most fundamental web applications, have become increasingly personalized in their provision of results, to the extent that they can also be considered recommender systems. 

One of the defining characteristics of recommender systems is personalization. Recommender systems are typically evaluated on their ability to provide items that satisfy the needs and interests of the end user. Such focus is entirely appropriate. Users would not make use of recommender systems if they believed such systems were not providing items that matched their interests. Still, it is also clear that, in many recommendation domains, the user for whom recommendations are generated is not the only stakeholder in the recommendation outcome. Other users, the providers of products, and even the system's own objectives may need to be considered when these perspectives differ from those of end users. Fairness and balance are important examples of system-level objectives, and these social-welfare-oriented goals may at times run counter to individual preferences. Sole focus on the end user hampers developers' ability to incorporate such objectives into recommendation algorithms and system designs. 

In addition, in many e-commerce retail settings, recommendation is viewed as a form of marketing and, as such, the economic considerations of the retailer will also enter into the recommendation function \citep{leavitt2006recommendation,pathak2010empirical}. A business may wish to highlight products that are more profitable or that are currently on sale, for example. Commercial recommender systems often use separate ``business rules'' functionality to integrate such items into the personalized recommendations generated through conventional means. Adding the retailer as a stakeholder allows such considerations to be integrated throughout the recommendation process.

We believe that, far from being special ``edge cases", these examples illustrate a more general point about recommendation, namely, that recommender systems serve multiple goals and that the purely user-centered approach found in most academic research does not allow all such goals to enter into their design and evaluation. What is needed is a shift in focus, a step back from a strict attention to the user to include the perspectives and utilities of multiple stakeholders.

In microeconomics, a similar shift occurred in the early part of the 21st century with the development of the theory of multisided platforms~\citep{rochet2003platform,evans_platform_2011}. Prior to that time, the traditional business model focused on a firm's ability to produce products and deliver them to customers at a price that could ensure profitability. This model was inadequate to explain Internet businesses such as search engines, that were giving their products away. Once multisided platform theory was developed, it enabled economists to look back at types of businesses that had existed for many years, such as credit card companies, shopping malls and stock exchanges, and recognize them as examples of multisided platforms as well~\citep{evans_matchmakers:_2016}. 

As noted above, when it comes to the study of personalized information access in the form of recommender systems, academic research has, with few exceptions, examined only a single side of these interactions. The stage was set historically by the first recommender systems implementations, which either operated on objects with no associated price (newsgroup posts~\citep{konstan1997grouplens}) or were external to any commerce associated with their recommendations (such as music, movie, and restaurant recommenders~\citep{shardanand1995social,breese1998empirical,burke1997findme}.) These systems brought users and products together, but they were not themselves party to any transactions. While academic research has largely concentrated on the user, commercial systems have regularly taken a broader view of recommendation objectives \citep{rodriguez2012multiple,nguyen2017multi}. There is, therefore, a gap between the complexity of real-world applications of recommender systems and those on which academic research has focused. 


The integration of the perspectives of multiple parties into recommendation generation and evaluation is the goal underlying the new sub-field of \textit{multistakeholder recommendation}~\citep{abdollahpouri_recommender_2017,soappaper,nguyen2017multi}. This article is intended to describe the current state of the art in multistakeholder recommendation research, to show some examples of current work in the area, and to outline research questions that should be addressed to support the demands of recommendation applications in environments where the perspectives of multiple parties are important. 

\section{Multistakeholder Recommendation}
\label{sec:MS_recommendation}

\noindent One important finding in the economics of multisided platforms is that different applications require different distributions of benefits or metrics of utility. In many multisided platforms, there is a ``subsidy side'' of the transaction where one set of parties uses the platform at a reduced cost or no cost. For example, users of the OpenTable restaurant reservation service do not pay to make reservations; instead, restaurants pay for each reservation made~\citep{evans_matchmakers:_2016}. Thus, Rochet and Tirole define a two-sided platform as one where the platform the volume of transactions varies with the consumer-side price, but not with the aggregate transaction price. If OpenTable customers are charged for their reservations, there will be fewer transactions, even if the platform lowers what it charges to restaurants. By contrast, a VAT tax can be divided up among the buyer and seller in different ways, and we expect this to have no effect on the market~\citep{rochet2004two}. 

The provision of recommendations does not necessarily entail a monetary transaction, but it is still worth considering users' interactions with a recommender system as a transaction. The user engages in some activity, for example visiting an e-commerce site, and recommendations are provided as a consequence of that interaction. Research has shown that consumers value such recommendations, thus we can attribute an associated utility for the user. Each transaction has the potential to impact others, such as manufacturers trying to sell their products, authors trying to sell their books, etc. A seller whose products are recommended in a given transaction will get some value from it, which could be considered the expected value of a recommendation turning into a sale or other similar benefit. There may not be a monetary cost, on any side, to using a recommender system, but instead there may be other kinds of costs: time investment, opportunity costs relative to other platforms, etc. 

\subsection{Definitions}

A recommender system can be abstractly defined in the following way. Let $U$ be a collection of users and their associated data. Let $I$ be a collection of items plus data, and let $R$ be a collection of data about the interactions between users and items, such as ratings, clicks, or purchases, with individual entries denoted as $r_{ui}$. We define a recommender system $S$ as a function $f$ that maps from a user, an item, and the interaction data to a score that will be used to rank items for recommendation to the user. 

\begin{equation}\label{eq:rec-fun}
    S: f(u,i,R)-> {\rm I\!R}
\end{equation}

We distinguish between a user-oriented recommender system and a multistakeholder recommender system by how we understand the output of this function. In a user-oriented recommender system, we interpret the output of the recommendation function as representing a prediction of the user's preference for the item. In other words, it will be comparable in some sense to the input interactions. Thus, among other properties, we expect a user-oriented recommendation function to be a type of mapping that preserves ordering properties of the input: 

\begin{equation}\label{eq:user-oriented}
    \forall i, j \in I: \mbox{if }r_{ui} \geq r_{uj} \rightarrow f(u, i, R) \geq f(u, j, R)
\end{equation}

This will not generally be true in a multistakeholder context. For example, a system might promote certain items in the interest of fairness towards item providers. In such a system, we do not interpret the output as strictly reflecting the user's preferences, and the condition of Equation~\ref{eq:user-oriented} is not expected to hold. Instead, we interpret $f$ merely as a scoring function implemented for the purposes of ranking. We analyze such a scoring function by examining how it combines the preferences of different stakeholders of which the user is one. 

Of course, as a practical matter, any recommender system is implemented by a single organization and so, trivially, would be intended to (only) represent its interests. However, when we analyze multistakeholder recommendation here, we are interested in developing a deeper understanding of how an organization might consider the perspectives of different parties -- the stakeholders -- in designing a recommender system. A multistakeholder recommender system is, therefore, one in which the ranking function cannot be understood as extrapolating from users' preferences to new items. Rather, the ranking function is best understood as representing the interests of multiple parties. In this paper, we are considering ranking functions that are designed to balance specific stakeholder needs, as opposed to market-based mechanisms that achieve such balance through economic means.

Although some applications might require a larger set, in this paper, we consider the following stakeholders: 

\begin{description}
    \item[Consumers \textit{C}:] The consumers are those who receive the recommendations. They are the individuals whose choice or search problems bring them to the platform, and who expect recommendations to satisfy those needs.
    \item [Providers \textit{P}:] The providers are those entities that supply or otherwise stand behind the recommended objects, and possibly gain utility from the consumer's choice. 
    \item [System \textit{S}:] The final category is the platform itself, which has created the recommender system in order to match consumers with providers and has some means of gaining benefit from successfully doing so. The platform may be a retailer, e-commerce site, broker, or other venue where consumers seek recommendations.
\end{description}

\subsection{Classes of multistakeholder applications}

\begin{figure}[tbh]
    \centering
    \includegraphics[width=3.5in]{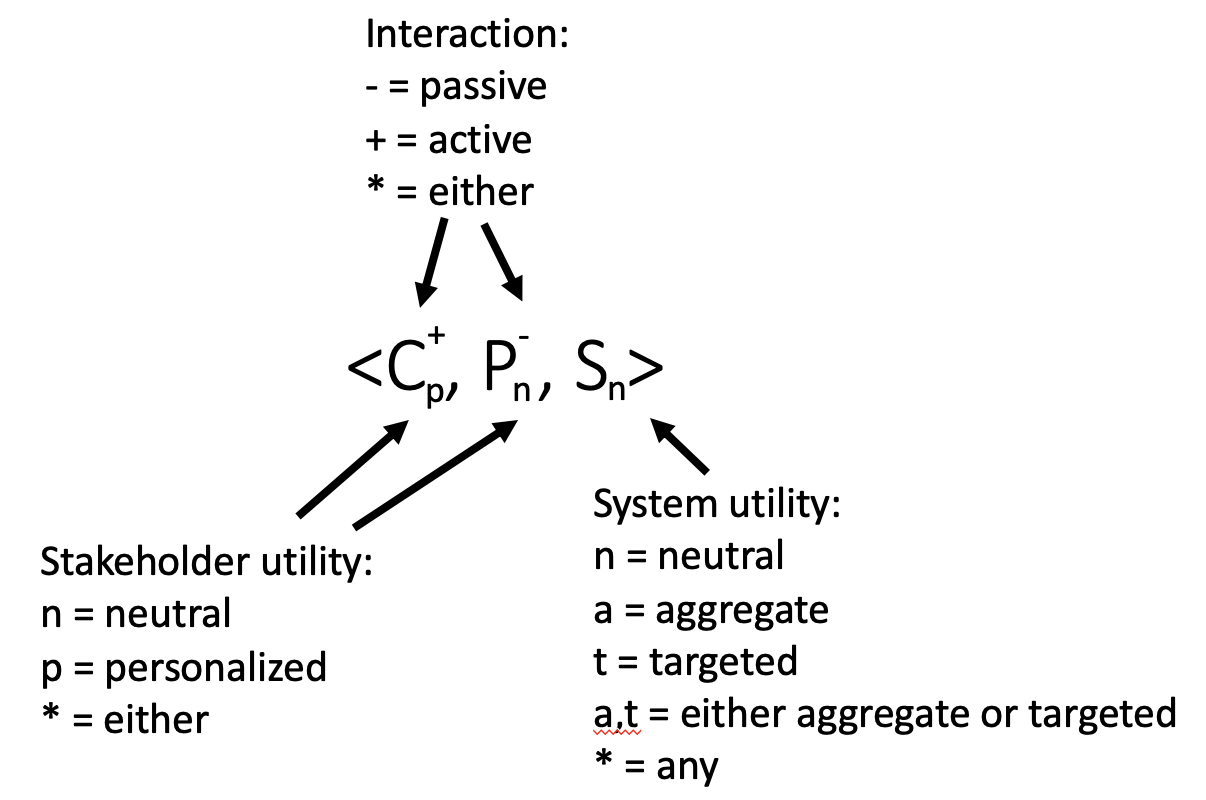}
    \caption{Stakeholder configuration notation.}
    \label{fig:notation}
\end{figure}

With these distinctions in mind, we can classify recommender systems based on how they support and incorporate the preferences of different stakeholders. See Figure~\ref{fig:notation} for a summary of the notation that we will use in this discussion. We represent each type of stakeholder (C, P, S) by an element in this triplet, and the requirements related to that stakeholder with associated subscripts and superscripts.

A system may support \textit{passive} or \textit{active} interaction (or both). A passive system provides recommendations unprompted by the user, such as ``Recommended for you'' items that are listed on many sites. We will use a superscript minus sign, as in $C^-$, to indicate the passive case. Active systems provide recommended items in response to some type of query or overt request from the user, denoted by the superscript plus sign: $C^+$. We extend this notion of active vs. passive to providers.

Another distinction has to do with the type of personalization an application supports. We can characterize personalization in terms of stakeholder preferences. If the system does not attempt to provide items that match stakeholder preferences, we say that the system is neutral, and use the subscript $n$. For example, an e-commerce site may not consider the provider when delivering recommendations. 

Typically, the system will attempt to satisfy the needs of at least one type of stakeholder by offering recommendations tailored (at least in part) to their preferences. This condition is denoted with the subscript $p$. When the stakeholder is the consumer, this is the most common personalized recommendation scenario. When there is personalization for both the consumer and the provider, we have reciprocal recommendation.

Table~\ref{tab:taxonomy} shows a taxonomy of possible stakeholder configurations when considering consumer and provider stakeholders only. We have dropped the $C_n$ configurations from this analysis as a system that is not personalized to end users does not generally qualify as a recommender system. Thus, the row contains only the passive vs active distinction. For providers, there is the additional possibility of neutral vs personalized, and therefore four columns. 

\begin{table}[tbh]
	\centering
\begin{tabular}{ |ll|p{1.75cm}|p{1.75cm}|p{1.75cm}|p{1.75cm}|}
\hline
\multicolumn{2}{|c|}{} & 
\multicolumn{2}{c|}{Passive} &
\multicolumn{2}{c|}{Active}\\
& & $P_n^-$ & $P_p^-$ & $P_n^+$ & $P_p^+$\\
\hline
Passive & $C_p^-$ & Standard recommendation     & Reciprocal recommendation & Paid placement & Display advertising \\
 \hline
Active  & $C_p^+$ & Personalized search & Reciprocal person search & Search advertising  & Reciprocal person search\\
 \hline
\end{tabular}
	\caption{$\langle C, P, S_n \rangle$ multistakeholder designs \label{tab:taxonomy}}
\end{table}

We can survey this space starting with the first row and column. This is the standard e-commerce recommendation scenario in which there is no attention paid to the provider side of the interaction. If the user formulates a query or other prompt to the system, we have a type of recommendation that personalizes search results, as is more and more common in search engines.

If the system is not neutral with respect to the provider, it must try to match consumers to provider preferences. Such designs are less common in research although perhaps not in practice. In the passive case (column 2), provider preferences are gathered implicitly through provider actions, for example, the acceptance of requests made by recommendation consumers -- such as an AirBnB host deciding whether or not to accept a guest. A system can learn provider preferences from such actions in the same way that it learns about the preferences of consumers, leading to the type of reciprocal recommendation designs discussed above.

In the active case, the provider can specify the type of consumer that is a desirable target. The neutral case (column 3) is one where the system does not attempt to personalize by learning differences between providers; it just matches users against providers' audience requirements. ``Promoted posts'' within social media sites are a good example -- these are recommendations given to users for whom they are a good fit, but only if the provider has indicated a specific interest in those consumers.

Where users are actively posing a query, we see familiar scenarios from search engine advertising. The placement of such ads takes place through a bidding process in which the providers' bids and the click likelihood are scored~\citep{edelman2007internet}. To the extent that click likelihood is calculated on a per-user basis, such systems could fall into either of the consumer active categories ($C_{*}^+$). 

A case that is arguably personalized with respect to the provider's utility is found in online display advertising. Here there is a form of reciprocation in that the ad should be appealing to the user and the user should be in the defined target audience. Real-time auctions are used in which providers are expected to quantify the utility they expect from a given ad placement~\citep{internetadvertisingyuan} and therefore the results for different providers can be different. Person-to-person search as in online dating sites also may incorporate reciprocal considerations. See Section~\ref{sec:datjob} below.

\subsection{System stakeholder}

Table~\ref{tab:taxonomy} only looks at cases where the design is neutral with respect to the recommendations it produces, not gaining or losing utility based on what results are produced. This is the classical model of how recommender systems operate. However, in many real-world contexts, the system may gain some utility when recommending items. When the system is included as a stakeholder, there is a third dimension to the taxonomy. 

Where the system does benefit from recommendation delivery, it may be in the form of a simple aggregate of the other stakeholders' utilities. For example, in many e-commerce settings, the system will get a commission for each sale, and such benefits can be considered together with personalization~\citep{nguyen2017multi}. Consistent with our notation, we will denote this case with the notation $S_a$ (aggregate) and the neutral case with $S_n$ (neutral).

Alternatively, the system may seek to tailor outcomes specifically to achieve particular objectives. For example, an educational site may view the recommendation of learning activities as a curricular decision and seek to have its recommendations fit a model of student growth and development. Its utility may, therefore, be more complex than a simple aggregation of those of the other stakeholders. This possibility of recommendations \textit{targeted} to system goals is designated with $S_t$.

\begin{figure}
    \centering
    \includegraphics[width=3.5in]{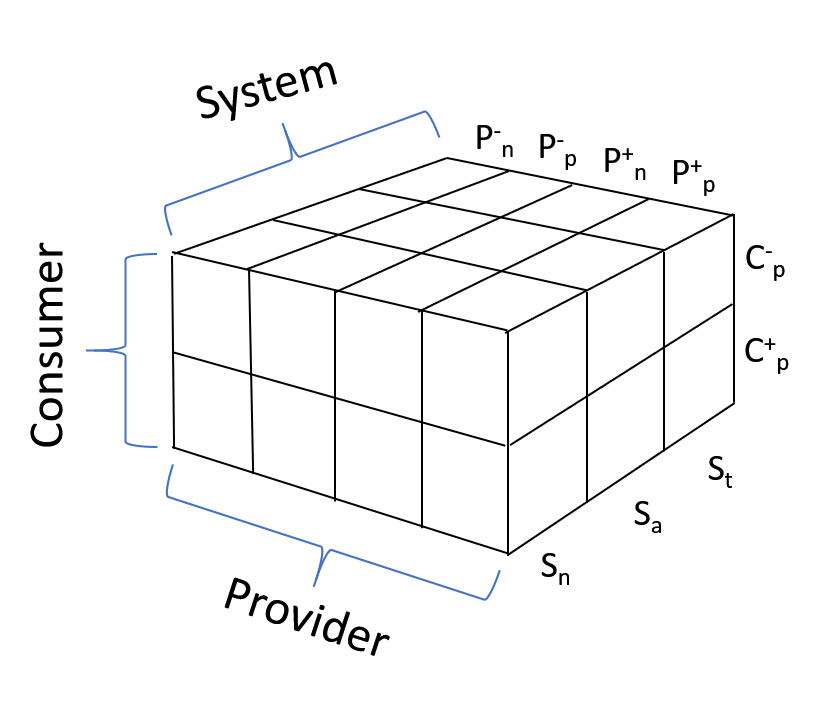}
    \caption{Multistakeholder design landscape}
    \label{fig:cube}
\end{figure}

This survey leaves us with a nuanced picture of the multistakeholder recommendation space, a three-dimensional extension of Table~\ref{tab:taxonomy} shown in Figure~\ref{fig:cube}. Each cell in this space can be defined by a triple $\langle C, P, S \rangle$, in which the type of interaction available to each stakeholder is specified. We have designs in which users are active or passive, that provide neutral and personalized results to providers, and allow for passive or active recommendation interactions, while the third dimension covers the different possibilities for system stakeholder. We will use the $\langle C, P, S \rangle$ triple notation throughout the discussion below to identify the area of the design space each application or area of research addresses. Many of these possibilities, especially the $S_t$ designs, have been rarely explored in the research literature.

\section{State of Current Research}

\noindent Multistakeholder recommendation brings together research in a number of areas within the recommender systems community and beyond: (1) in economics, the areas of multisided platforms and fair division; (2) the growing interest in multiple objectives for recommender systems, including such concerns as fairness, diversity, and novelty; and, (3) the application of personalization to matching problems.

\subsection{Economic foundations}

\noindent The study of the multisided business model was crystallized in the work of Rochet and Tirole \citep{rochet2003platform}on what they termed ``two-sided markets.'' Economists now recognize that such contexts are often multisided, rather than two-sided, and that ``multisided-ness'' is a property of particular business platforms, rather than a market as a whole~\citep{evans_matchmakers:_2016}. Prior to their work, the traditional business model focused on a firm's ability to produce products and deliver them to customers at a price that would ensure profitability. This model was inadequate to explain Internet businesses such as search engines that were giving their products away. 

Multisided platforms bring together market participants, and create value by lowering their transaction and search costs~\citep{evans_platform_2011,evans_matchmakers:_2016}. Many of today's recommender systems are embedded in multisided platforms and hence require a multisided approach. The business models of multisided platforms are quite diverse, which means it is difficult to generalize about multistakeholder recommendation as well. A key element of the success of a multisided platform is the ability to attract and retain participants from all sides of the business, and so developers of such platforms must model and evaluate the utility of the system for all stakeholders. 

The theory of just division of resources has a long intellectual tradition going back to Aristotle's well-known dictum that ``Equals should be treated equally.'' Economists have invested significant effort into understanding and operationalizing this concept and other related ideas. See~\citep{moulin2004fair} for a survey. In recommendation and personalization, we find ourselves on the other side of Aristotle's formulation: all users are assumed unequal and unequal treatment is the goal, but we expect this treatment to be consistent with diverse individual preferences. Some aspects of this problem have been studied under the subject of social choice theory~\citep{arrow2010handbook}. However, there is not a straightforward adaptation of these classical economic ideas to recommendation applications as the preferences of users may interact only indirectly and in subtle ways. For example, if a music player recommends a hit song to user A, this will not in any way impact its desirability or availability to user B. On the other hand, if the job recommender system recommends an appealing job to user A, it may well have an impact on the utility of the same recommendation to user B who could potentially face an increased competitive environment if she seeks the same position.

\subsection{Multi-objective recommendation}
Multistakeholder recommendation is an extension of recent efforts to expand the considerations involved in recommender system evaluation beyond simple measurements of accuracy. There is a large body of recent work on incorporating diversity, novelty, long tail promotion and other metrics as additional objectives for recommendation generation and evaluation. See, for example, \citep{abdollahpouri2017controlling,diversitySmyth,diversityziegler,noveltydiversityvargas,jannach2016recommendations}. There is also a growing body of work on combining multiple objectives using constraint optimization techniques, including linear programming. See, for example, \citep{jambor_optimizing_2010,agarwal_click_2011,svore_learning_2011,rodriguez_multiple_2012,jiang_optimization_2012,agarwal_personalized_2012}. These techniques provide a way to limit the expected loss on one metric (typically accuracy) while optimizing for another, such as diversity. The complexity of these approaches increases exponentially as more constraints are considered, making them a poor fit for the general multistakeholder case. Also, for the most part, multi-objective recommendation research concentrates on maximizing multiple objectives for a single stakeholder, the end user. 

Another area of recommendation that explicitly takes a multi-objective perspective is the area of health and lifestyle recommendation. Multiple objectives arise in this area because users' short-term preferences and their long-term well-being may be in conflict~\citep{lin2011motivate,ponce2015quefaire}. In such systems, it is important not to recommend items that are too distant from the user's preferences -- even if they would maximize health. The goal to be persuasive requires that the user's immediate context and preferences be honored.

Fairness is an example of a system consideration that lies outside the strict optimization of an individual user's personalized results. Therefore, recent research efforts on fairness in recommendation are also relevant to this work~\citep{lee2014fairness,burke2017balanced,burke_multisided_2017,DBLP:conf/recsys/KamishimaAAS14,kamisha-akaho-fatrec-2017,yao_huang_fatml-2017}. The multistakeholder framework provides a natural ``home'' for such system-level considerations, which are otherwise difficult to integrate into recommendation. See Section~\ref{sec:far} for a more in-depth discussion.

\subsection{Personalization for matching} 

\noindent The concept of multiple stakeholders in recommender systems is suggested in a number of prior research works that combine personalization and matching. The earliest work on two-sided matching problems~\citep{roth1992two} assumes two sets of individuals, each of which has a preference ordering over possible matches with the other set. The task to make a stable assignment has been shown to have an $O(n^2)$ solution. This formulation has some similarities to reciprocal recommendation. However, it assumes that all assignments are made at the same time, and that all matchings are exclusive. These conditions are rarely met in recommendation contexts, although extensions to this problem formulation have been developed that relax some of these assumptions in online advertising contexts~\citep{bateni2016fair}.

Researchers on reciprocal recommendation have looked at bi-lateral considerations to ensure that a recommendation is acceptable to both parties in the transaction~\citep{reciprocal}. See Section~\ref{sec:datjob} for a detailed discussion. A classical example is on-line dating in which both parties must be interested in order for a match to be successful~\citep{reciprocaldating}. Other reciprocal recommendation domains include job seeking~\citep{rodriguez_multiple_2012}, peer-to-peer ``sharing economy'' recommendation (such as AirBnB, Uber and others), on-line advertising \citep{targetadvertisingbiding}, and scientific collaboration~\citep{lopes2010collaboration,tang2012cross}.

The field of computational advertising has given considerable attention to balancing personalization with multistakeholder concerns. Auctions, a well-es\-tab\-lished technique for balancing the concerns of multiple agents in a competitive environment, are widely used both for search and display advertising \citep{internetadvertisingyuan,mehta2007adwords}. However, the real-time nature of these applications and the huge potential user base makes recommender-style personalization computationally impractical. 

\section{Examples}

In the following sections, we introduce three applications of multistakeholder recommendation. These examples show some of the variety of contexts in which this concept can be applied. 

Section~\ref{sec:people} looks what is perhaps the earliest application area for multistakeholder recommendation -- reciprocal recommendation. In reciprocal recommendation, the recommender system matches users with other users, thus collapsing the distinction between consumers and providers: the consumers of recommendations are also the individuals who might be recommended to others. Depending on the application, reciprocal recommendation designs occupy the section of taxonomy represented by $\langle C^*_p, P^*_p, S_n \rangle$, where any combination of active or passive interaction on the part of users might be part of the design. 

Section~\ref{sec:value} examines the broad class of $\langle C^*_p, P^*_*, S_{a,t} \rangle$ designs where the system's interest is enhancing profit or economic value related to recommendations that are produced. It looks specifically at a real-world example of a commission-based system where recommendations can serve as unpaid advertising to the provider. In this case, the economic viability of the recommendation platform depends on the ability to prioritize items that are likely to generate commissions. 

The final example in Section~\ref{sec:far} examines the problem of fairness in recommendation. Fairness is inherently a multistakeholder concept. If the only consideration in recommendation generation is matching individual user's known preferences, then the question of whether recommendations are fair does not arise. Fairness is therefore a quintessential $S_t$ system-level concern, not reducible to the problem of maximizing aggregate utility for either consumers or providers. Most fairness research takes the form of $\langle C^-_p, P^-_n, S_t \rangle$, a variant of the standard consumer-oriented recommendation scenario, but one where fairness of one type or another is important.

\section{Example: People Recommendation}
\label{sec:people}
    People recommendation is based on the notion of social matching~\citep{terveen05matching}, as discussed above. The fact that the recommended entity is a person yields additional reciprocity and requires additional considerations of trust, privacy, reputation, and personal attraction. The symmetry between the individuals means that these recommenders will have always both $C_p$ and $P_p$ aspects. They may also have different types of system objectives as discussed below. 

\subsection{Reciprocal Recommenders}
\label{sec:datjob}

A reciprocal recommender as defined in \citep{Pizzato:UMUAI:2012} is a people-to-people recommender system in which the preferences of both sides of the recommendation should be taken into account (the user receiving the recommendation and the user being recommended). 
Online dating recommenders as well as employment matching (talent recommenders and job recommenders) are prime examples of reciprocal recommenders, given that no successful match or recommendation of a date or job placement will occur without both parties agreeing to it. Reciprocal recommenders differ from traditional non-reciprocal recommenders in a number of ways. Table~\ref{t:diff:tradandreciprocal:recsys} shows some of these main differences as described in the study done in \citep{Pizzato:UMUAI:2012}.

\begin{table}[t]
\caption{Main differences between traditional recommenders and reciprocal recommenders}
\label{t:diff:tradandreciprocal:recsys}
\centering
\begin{tabular}{|p{.4\linewidth}|p{.4\linewidth}|}
\hline \bf{Traditional Recommenders} & \bf{Reciprocal Recommender} \\

\hline 	Success is determined solely by the user seeking the recommendation &

	   	Success is determined by both subject and object of the recommendation \\

\hline 	Users have no reason to provide detailed explicit user profiles. &

		Users expect to provide detailed self-profiles. 
		Explicit profiles and preferences are often inaccurate. \\

\hline	Satisfied users are likely to return for more recommendations. 
		Better recommendations means more engagement; &

		Users may leave the system after a successful recommendation; 
		Better recommendations might mean less engagement; \\

\hline	Items can get recommended even when stocks are low; 
		Information items are always available; &
		
		It is important not to overwhelm users by recommending popular users to others too often; \\

\hline

\end{tabular}
\end{table}

As the table indicates, user behaviour is highly dependent on whether a domain is reciprocal or not. The success of a traditional book recommender is dependent only on the person receiving the recommendation. On the other hand, in a reciprocal domain such as online dating, the user receiving the recommendation knows that the contact suggested by the recommender is only going to be successful if the other user also agrees to the contact. The users of reciprocal domains can act either proactively by taking the initiative to connect with other users or simply being reactive and waiting for contact, in effect choosing either the consumer or provider role. 

The ``free-rider'' problem is well-known in the recommender systems and other online rating systems that depend on explicit user profiles. Users may benefit from others' contribution without adding their own and may lack an incentive to build up their profiles. In contrast, for reciprocal recommendation, profiles have a communicative role in providing reasons to propose or accept a contact, a clear need and benefit to provide rich user profiles.\footnote{For this reason, however, profiles are sometimes inaccurate~\citep{okcupid2010lies}, something that reciprocal recommenders need to account for.} Conversely, however, user profile data consisting of interactions with items may be sparser than in traditional recommendation contexts. In settings such as online dating and job recommendation, the task is often to find a suitable match quickly and exit the market. Users may only require a few interactions to achieve this goal, as opposed to consumption-oriented contexts where a user might rate dozens of books or hundreds of music tracks.

Because a successful recommendation in a reciprocal domain means that the user is likely to leave the system, conflicting incentives are created for platform owners. They want to have a profitable business by having repeated users, even though the best user experience would be for each user to instantly find a match (a successful date/partner) and never return.

Another consideration important in reciprocal recommendation is that, unlike other recommendation settings, the system's utility is not necessarily increasing function of the volume of recommendations. For instance, imagine if a highly qualified person is recommended to every single job position that they are fit to hold. This person is likely to be burdened by the amount of contact and might leave the website. A similar situation can occur for popular users in a dating website. These users are important as they represent the best of each of these services, but they can easily be overwhelmed by the interest of other users. Such a user should only be recommended to others when the recommender is highly confident that they will reciprocate. 

Many systems where people are in both sides of the recommendation process benefit from reciprocity even when reciprocity is not required in the system. In some of these domains such as Twitter, relationships have shown to be stronger, with a lower likelihood on one breaking a social link, when both users follow each other \citep{xu:CSCW:2013,Kwak:2011:FOR,Kivran-Swaine:CHI:2011}. See Section~\ref{sec:interesting-people} below.

\subsubsection{Online Dating Recommendation}
Reciprocal recommenders perfectly fit the problem of job recommenders and online dating recommenders since these platforms define success in reciprocal terms (i.e. if one side of the recommendation disagrees with a match then the recommendation is not successful). That is not to say that these systems (and their recommender systems) cannot exploit unbalanced market forces that would make harder to satisfy a user with a large demand and small supply (e.g. a highly skilled job seeker in a highly-demanded profession). 

Online dating websites focus on recommending people that one might like to date. Some websites also focus on friendships, therefore recommending people one does not know yet. Dating websites could be built on existing relationships and many have tried using existing social networks for this. However, one of the reasons people go to dating websites because their existing social connections do not provide them with the people with whom they are interested in connecting romantically. 

RECON \citep{Pizzato.etal:RecSys:2010} was the first recommender system to exploit the benefits of reciprocity in the online dating context. This system works by calculating a compatibility score between users and recommending people to people who have higher reciprocal compatibility scores. A number of studies followed this, including designs that focus on improving the cold-start problem of reciprocal recommenders \citep{Akehurst.etal:IJCAI:2011,Yu_et_al:ASONAM:2016}. 

Building on this work, Li and Li \citep{Li.Li:CIKM:2012} proposed MEET, a generalised reciprocal framework in order to integrate a number of aspects related to the reciprocal domain and in particular in online dating. MEET uses a bipartite graph that represent the mutual interest among a set of users to another set of users (men and women, in a heterosexual dating network). By creating subgraphs, it is able to perform graph inference and obtain a list of recommendations that is ranked based on mutual interests and filtered for users who exceed a certain availability budget. 

Xia et al. \citep{Xia_et_al:SNAM:2016} proposed and compared a number of online dating recommenders including reciprocal content-based, memory-based and model-based collaborative filtering and found that memory-based methods outperform model-based models for female users who tend to have the largest sparsity in their interaction matrix. 

Goswami et al. \citep{Goswami_et_al:ICML:2014} discusses reciprocal recommenders in more general terms as a two-sided market and proposes a two-layer architecture for recommendation ranking that looks at the preferences of both sides of the market. Alanazi and Bain \citep{Alanazi_Bain:IWMLRC:2016} developed a reciprocal recommender using hidden Markov models.

\subsection{People Recommendation on Social Media}
At the core of social media are individual relationships which serve as a fertile ground for recommendation. The underlying social network of a social media website -- explicit through articulated connection or implicit via shared interests or goals -- drives diffusion and engagement as well as key features such as news feeds and photo streams. The network's size is often considered a key metric of a social site's success. Recommendations of people on social media sites therefore play a key role in their success and have become ubiquitous~\citep{guy18people}.

Three primary techniques are used for people recommendation on social media:

\begin{itemize}
    \item Graph-based techniques that consider the graph representation of the network and apply link prediction algorithms, such as shortest paths, PageRank, and clustering, on top of it. The edges of the graph may carry different semantics according to the specific site and its underlying network.
    \item Interaction-based techniques that consider different types of user interaction with content, which are widespread and diverse on social media. These include bookmarking, commenting, tagging, sharing, `liking', joining, and more.
    \item Content-based techniques that use the actual content, typically tying it to its authors. Different text and image processing techniques, such as language modeling, syntactic parsing, word embedding, and object detection are applied. 
\end{itemize}

The literature on people recommendation has seen a substantial growth in recent years. Recommendation types can be grouped based on the type of relationship and its intended duration. Specifically, three types of relationships -- familiarity, similarity,  and interest  -- differentiate between the different networks. Recommendations also differ on whether the action suggested results in a permanent or temporary / ad-hoc relationship. In the rest of this section, we will discuss the common aspects and differences between all six types of recommendations, according to their relationship type and its intended action. 

\subsubsection{Recommending Familiar People}
The most fundamental scenario of people recommendation on social media suggests familiar people for a long-term (permanent) connection, namely the recommendation of people to connect with on social network sites (SNSs), whose primary type of connection is symmetric (confirmed), such as Facebook and LinkedIn. This type of recommendation benefits both sides and reciprocity is its main characteristic. As a result, the person who receives the recommendation knows the other party (the recommended person) would have to confirm the connection and this party's anticipated reaction plays a key role in the decision making process leading to accepting or ignoring the recommendation. The permanent type of the connection also entails high weight -- accepting such a recommendation may lead to a multi-year connection with another person on the SNS, which would involve receiving updates, news, photos, posts, and other types of information over a long period of time. 

Widgets that proposed ``people you may know'' started to appear on leading SNSs at the end of the previous decade. Early work conducted on symmetric social networks within the enterprise, showed the benefit of aggregating multiple signals for recommendation~\citep{chen09make} and indicated a dramatic effect on the number of connections on the site~\citep{guy09dyk}. It was also shown that providing evidence for a person's recommendation, such as the joint documents they have with the individual who receives the recommendation, helps making the latter feel more conformable accepting the recommendation and triggering the invitation to connect. 

Two interesting followup studies were conducted by Guy et al. ~(\citeyear{guy09dyk}), inspecting longer term effects. In the first, people recommendation were shown to increase engagement and retention rates on enterprise SNS when new users were introduced with people recommendations~\citep{freyne09increasing}. In this scenario, it was shown that the most effective recommendation ranking was by activity on the site rather than by the total weight of connection signals to the target user. Apparently, recommending active people has a special impact when trying to engage new users. The second followup study focused on the network effects of the provided recommendations~\citep{daly10effects}. It was shown that different people recommendation algorithms render different network structures, as reflected in a variety of social network analysis metrics, such as betweenness centrality. Incorporating a global criterion into the recommendation strategy moves this type of recommendation from the $S_a$ category, where the system is only interested in increasing the number of connections made, to the $S_t$ category, where the system is aiming to optimize a target criterion independent of individual outcomes. As noted in Section~\ref{sec:MS_recommendation}, these types of designs have not received much attention in the literature. 

The ad-hoc scenario of familiar people recommendation is especially common on mobile devices, which provide location awareness and other types of contextual properties. Henceforth, recommending two people who are already familiar with each other, in a specific context, such as when shopping in the same mall or when going to same concert, becomes a handy use case. Yet, privacy considerations must be respected, since this type of recommendation requires exposure of one's location to others. A different example of the ad-hoc scenario for familiar people is demonstrated by Cluestr, which suggested contacts that can be addressed as a group, such as co-workers, family members, or friends, to support communication in a closed group and save time in the group initiation process~\citep{grob09cluestr}. 

\subsubsection{Recommending Similar People}
The second relationship type involves similar individuals, where naturally the number of potential candidates is the largest. Similarity signals are abundant on social media, due to the rich types of content and user associations with it~\citep{guy16like}, and generally map to similarity in places, things, and people~\citep{guy10same}. Such signals can also be used to match-make strangers in the enterprise~\citep{guy11strangers}. This example of a permanent recommendation task is considered an exploratory recommendation use case, where the hit rate is expected to be lower than other types of recommendations, since the targets are unfamiliar. Yet, the value for a successful recommendation is much higher: getting to know a new individual who has similar interests and with whom a long-term relationship may be built will increase one's social capital. A stranger recommender will need to subtract familiarity signals from the similarity signals to detect distant co-workers who may share common interests with the target employee. Among other scenarios for permanent recommendation of similar people are the dating and employee/employer recommendations described in Section~\ref{sec:datjob}, when applied in the context of social media websites.

The scenarios for ad-hoc recommendation of similar people are plentiful. A prominent example is the recommendation of similar people at a conference. Many resources are invested in organizing and traveling to the conference, where networking and meeting new individuals is one of the key goals. For example, in the academic research domain, such meetings can assist in receiving help or advice with ongoing projects, acquiring new opportunities, finding potential partners, and discovering research directions. Having said that, such meetings typically occur on a rather arbitrary basis. Find \& Connect~\citep{chin12find} used physical proximity via RFID badges to recommend new contacts at a conference, while SPARP~\citep{xia14socially} combined weak ties with similar personal traits to generate recommendations. Both systems were used and evaluated at real venues. 

Two additional scenarios are the suggestion of guests for invitation to an event and the recommendation of activity partners. For the first, both relevance of the event to the guest and geographical proximity were considered and evaluation was performed using Foursquare data in London~\citep{saez12ads}. 
Activity partner recommendation was suggested as an accompanying scenario to item recommendation, such as for a movie or a dinner~\citep{tu15activity}. The recommendation algorithm considered the user's similarity of interests to the target as well as their probability to like the item itself. Evaluation indicated that this kind of partner recommendation can improve the success of item recommendation. 

\subsubsection{Recommending Interesting People}\label{sec:interesting-people}
Interest plays a key role in the user-item space for typical item recommendation, yet it also occurs as a relationship type between two users~\citep{jacovi11digital}. This relationship type is often associated with a ``follow'' recommendation in asymmetric networks, where a user can connect to another party without their need to approve or reciprocate~\citep{gupta13wtf}. With reciprocity removed, this type of recommendation more closely resembles the standard $\langle C^-_p, P^-_n, S_n \rangle$ single stakeholder scenario. 

For ad-hoc recommendation of interesting people, a salient scenario that received substantial attention in recent years is the recommendation of people to mention on a tweet or other types of posts. For tweets, which are limited in text length, the selection of individuals to mention is also limited and associated with a concrete cost. On the other hand, mentions play a key role in information dissemination, as mentioned individuals often re-post to their own audience, thus entailing a potential reciprocal relation. From a marketing perspective, the recommendation should take into account special types of signals in addition to the relationship of the two users, such as the content relevance and the response likelihood. In addition, the recommended user's popularity and influence can also play a role~\citep{wang13whom,jacovi14reputation}. As a result, typical solutions combine graph-based, interaction-based, and content-based methods to produce the ultimate list of recommendations~\citep{wang13whom,gong15who,huang17mention}.

\subsection{Summary}
The examples of people recommendation we provided in this section show that a traditional user-oriented recommender system $\langle C^-_p, P^-_n, S_n \rangle$ might not deliver the best recommendations for those scenarios. Problems like reciprocal recommendation requiring $P^-_p$ or $P^+_p$ designs, where the needs and interests of both parties (the person who is being recommended and the person who receives the recommendation) can be taken into account. Moreover, in some reciprocal recommendation applications like online dating, the recommender system may also need to attend to the differences in desirability level of different users and try not to overwhelm certain users while not giving enough exposure to others. Thus, the distribution of recommendations becomes important. In addition, in some situations, the system might also have goals regarding the overall structure of the network constructed by connecting people with each other. For example, websites like Twitter, may want their network of users have certain properties (betweenness degree, centrality etc.) and would therefore, need a $\langle C^*_p, P^*_n, S_t \rangle$ design to  deliver recommendations targeted towards their desired network structure.

\section{Example: Value-aware Recommendation}
   \label{sec:value}

The literature in the field of recommender systems, as mentioned in Section \ref{sec:introduction}, mainly focuses on the consumer perspective with the system being neutral regarding what items get recommended: the common $\langle C^-_p, P^-_n, S_n \rangle$ design. The goal of most research efforts is therefore to design algorithms and systems that aim to provide value for the consumer in some form, e.g., by avoiding information overload or helping the consumer to discover new items. Even then, in many cases, researchers tend to abstract away from real-world consumer value metrics, such as consumer surplus or satisfaction, and focus on optimizing more general algorithmic metrics such as RMSE, NDCG, or precision and recall.

The underlying implicit assumption here is that recommending only assumedly relevant items to the user will also have a positive impact on the value for the provider or the platform. In fact, a number of studies support this hypothesis and show that providing personalized recommendations that are optimized to match the user's preferences lead to increased business value, e.g., in terms of increased sales or click-through rates \citep{Garcin:2014:OOE:2645710.2645745,DBLP:conf/pkdd/KirshenbaumFD12,DBLP:conf/recsys/JannachH09}; and, vice-versa, that unexpected or irrelevant recommendations can lead to a decreased quality perception and trust by consumers \citep{Chau:2013:EEM:2747904.2748214,fitzsimons2004reactance}. 

\subsection{The business value of recommender systems}
Generally, there are different ways in which a recommender can create value for a provider or a platform. Considering the system or platform stakeholder, e.g., an online retailer or streaming media site, a recommendation service on the site can serve multiple purposes. It can, for example, lead to more overall sales as mentioned above, it can more indirectly lead to increased consumer engagement and loyalty, and it can even be a competitive factor when other actors on the market do not have a recommendation service \citep{jannach2016recommendations}. Thus, a system might treat the utility of the recommender system as being a purely aggregate function of the utility delivered to end users: the $S_a$ condition.

While factors like increased consumer engagement often considered to lead to indirect business value, e.g., in terms of an increased number of monthly re-subscriptions, recommendations can also be used to positively impact the business in a more direct way. Specifically, recommenders can be implemented as a tool that steers consumer demand, e.g., by promoting certain items. The particular goal in that context can be to drive demand in a direction that maximizes the platform's short-term or long-term profit, 
while also maintaining an acceptable level of consumer utility. A system of this type would fall in the $S_t$ category.

\subsubsection{The need for a balanced approach}\label{sec:balanced}
Simply recommending those items with the highest profit for the platform is probably in almost all cases not the optimal strategy, at least not in the long run, as consumers might start to distrust a recommendation service when its suggestions are not considered useful. Generally, we can hypothesize that in many domains there is a trade-off between suggesting items that are the most profitable for the platform and suggesting those that are considered the most relevant for the user.

However, an additional intuitive assumption in that context might also be that a recommender is still effective in steering consumer demand if the order of the suggestions is not strictly determined by the assumed relevance for the consumer, but also takes profitability considerations into account.

\begin{figure}[h!t]
    \centering
    \includegraphics[trim=80pt 125pt 70pt 115pt, clip, width=0.66\textwidth]{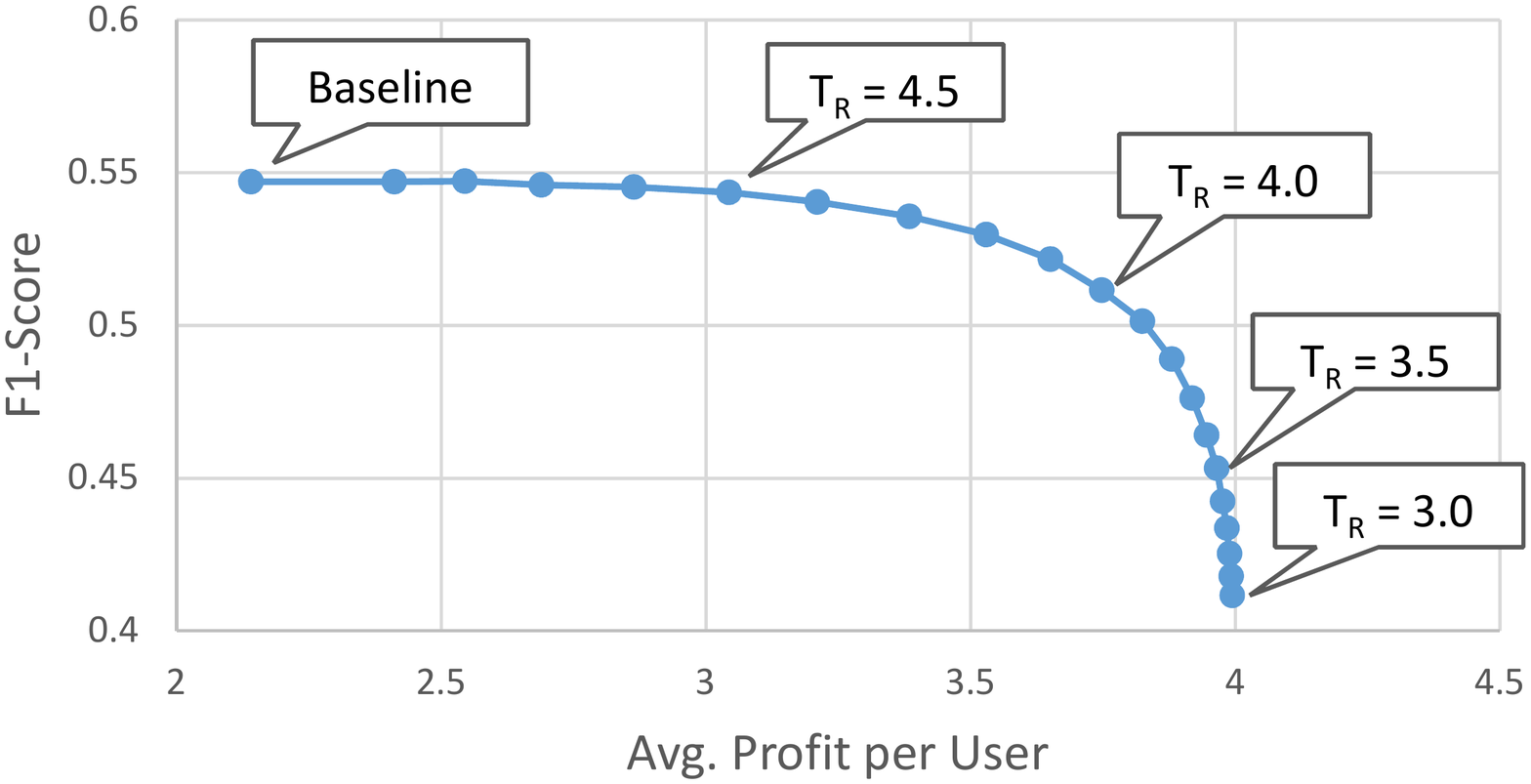}
    \caption{Profit-relevance tradeoff (guaranteed purchase), from \citep{JannachAdomaviciusVAMS2017}.}
    \label{fig:numerical-experiment}
\end{figure}

\begin{figure}[h!t]
    \centering
    \includegraphics[trim=70pt 295pt 70pt 330pt, clip, width=0.66\textwidth]{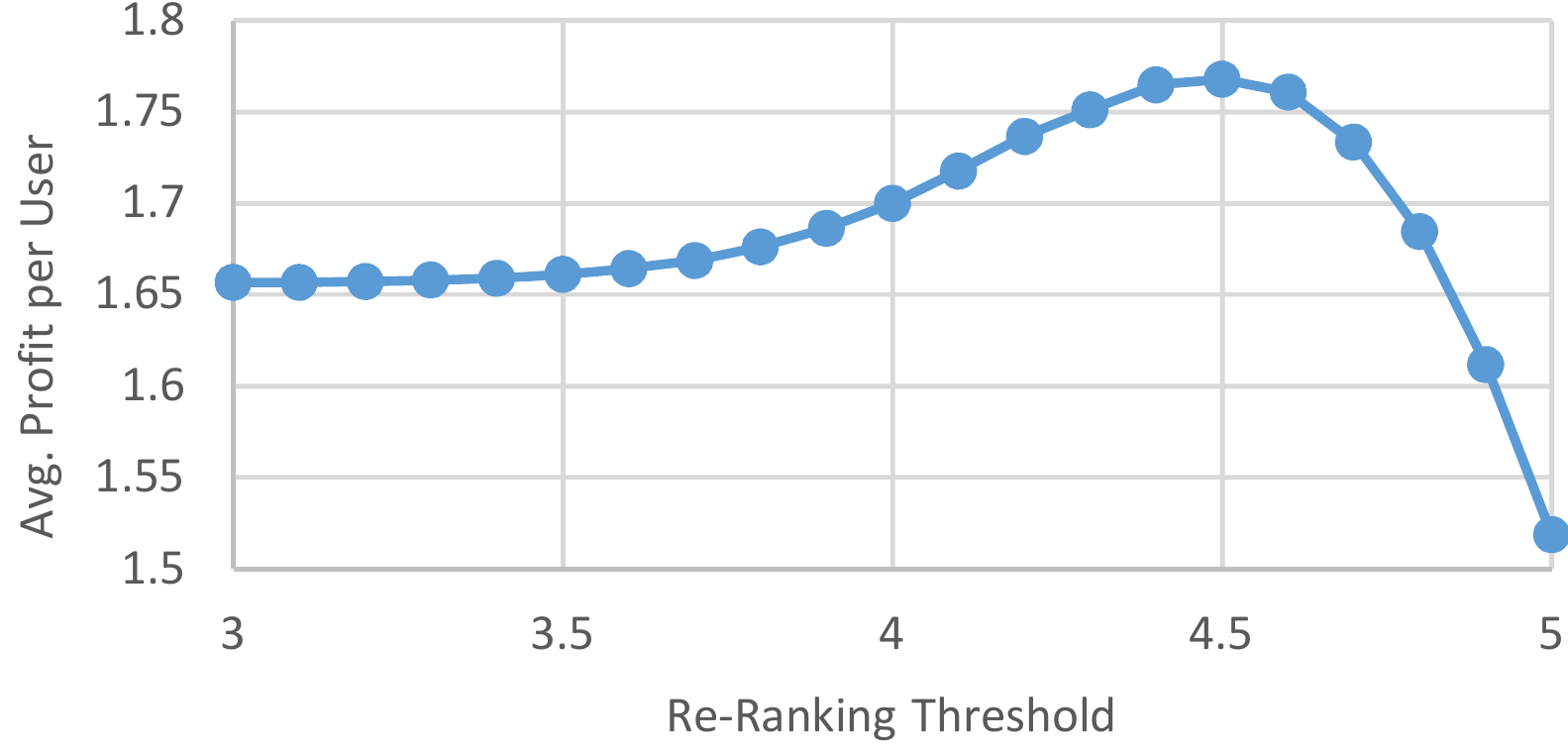}
\caption{Profit optimization (relevance-based purchase), from \citep{JannachAdomaviciusVAMS2017}.}
    \label{fig:profit-modeling}
\end{figure}

As an illustration for such a potential trade-off, let us consider the outcomes of two simulation experiments shown in Figure \ref{fig:numerical-experiment} and Figure \ref{fig:profit-modeling} \citep{JannachAdomaviciusVAMS2017}. To conduct the simulations, artificial profitability data were added to the items of the well-known MovieLens 1M dataset. Specifically, each movie in the dataset was assigned a profitability value taken from a Gaussian distribution, with a mean value of \$2, a standard deivation of \$1, and upper and lower threshold values of \$4 and \$0, respectively.

In the first simulation experiment, we adopted 
the greedy item re-ranking scheme from \citep{DBLP:journals/tkde/AdomaviciusK12} to maximize the profit of the top-10 recommended items. Technically, the method takes the relevance-optimized list of an underlying recommendation algorithm, in our case a matrix factorization technique, and pushes the items with the highest profit up the list, as long as their predicted relevance surpasses some minimum threshold $T_R$. Figure \ref{fig:numerical-experiment} shows that without the consideration of the profit, in the case of providing top-N item recommendations by explicitly considering only item relevance (i.e., ``Baseline'' in Figure \ref{fig:numerical-experiment}), the average profit per user is slightly higher than \$2. However, when we take profitability information into account, even if we restrict ourselves to recommending items for which the expected relevance (rating) is higher than 4.5 (on a 1-to-5 scale), we can increase the profit by more than 50\% to over \$3, with only a very limited loss on the overall recommendation accuracy (as shown in terms of the F1 measure on the Y-axis).

In this first simulation, the underlying assumption is that consumers will always pick one of the recommended items. In some domains, however, it might happen that in case the recommendations are unsuited -- due to limited relevance or generally poor quality -- the consumer might decide not to pick any of the recommendations at all, leading to a profit of 0. In the second simulation we therefore introduced an exponential decay function that models a decreasing purchase probability when the predicted relevance is lower.
The decay function was modeled in a way that we assumed a 10\% probability that an item was purchased when it had a predicted rating of 5, which leads to a high expectation of a purchase if all ten items in the recommendation list have the highest possible rating. The decreasing acceptance probability $p$ of an item with a predicted rating $r$ was then modeled as follows.

\begin{equation}
p = \frac{1}{10} \times e^{\alpha*(5-r)}
\end{equation}
where $\alpha$ is a parameter that we set to -1.5 for this simulation.




When no re-ranking is done in this simulation the average profit per user is at about 1.30. Figure \ref{fig:profit-modeling} shows the effects of applying the profit-aware re-ranking scheme. According to the simulation, the optimal threshold is at about 4.5. When further lowering the threshold, even more profitable items become available to be placed in the top-10 list. However, the higher profit cannot compensate the rapidly decreasing purchase probabilities for such items. When the threshold is set higher than 4.5, the problem exists that there are not many (high-profit) items are left that can be moved up the list. Nonetheless, the average profit per user is still higher than the baseline profit of 1.30.

The simulations in the two experiments are based on a comparably simple re-ranking scheme. In reality, more elaborate techniques can be applied which in addition takes the price sensitivity of the individual user into account. In \citep{JugovacJannachLerche2017eswa}, for example, a corresponding generic optimization-based re-ranking method is described, which is designed to balance multiple, possibly conflicting optimization goals and to consider consumer-individual preferences regarding this balance.

\subsection{Review of Selected Profit-Aware Approaches}\label{sec:review-profit}
In the literature, a variety of methodological approaches of different complexities were explored to incorporate profit information into recommenders and to balance relevance and profitability.

Considering not only the consumer preferences but also the profitability for the seller, as discussed in the example above, was in the focus in \citep{Chen20081032}. In this work, the authors compared different recommendation strategies that combine general and customer-individual purchase probabilities of the items with profitability information on synthetic data. Their simulations, like ours above, indicate that higher overall profitability can be achieved without a loss of accuracy for personalized recommendations. Focusing too much on profitability, however, leads to an accuracy degradation.

Going beyond the comparably simple model of \citep{Chen20081032}, Wang and Wu \citep{Wang20097299} framed the selection of products for customers as a constrained optimization problem. The constraints in the model ensure that the recommended products match the customer's preferences and their assumed budgets. This work therefore also considers the user's price sensitivities. Different optimization goals can be configured which either maximize the profit for the seller or lead to a win-win situation where seller profits and customer value are balanced. Alternative approaches that model the recommendation problem as a mathematical optimization task were later on put forward in
\citep{DBLP:journals/corr/abs-0908-3633,ValuePick2010,Hammar:2013:UMC:2507157.2507169,Azaria:2013:MRS:2507157.2507162}.
The model proposed in \citep{DBLP:journals/corr/abs-0908-3633}, for example, includes the concept of \emph{trust}, assuming that a consumer will continue to make purchases as long as the system is able to predict their preferences to a certain extent. The proposed work unfortunately remained on a theoretical level and it is in particular unclear to what extent the assumed trust model is realistic.

Lu et al.~\citep{Lu:2014:SMM:2733085.2733086} take yet another set of factors into account in the revenue model of their optimization-based approach, including prices, saturation effects and competition effects. A specific aspect of their work is that they optimize the model over a finite time horizon, where the adoption probability at each time step can depend on different factors such as the previous purchases by the consumer in the same class, the number of times a certain item was already recommended, or the current price of the item and the individual consumer's willingness-to-pay (WTP) for it. Given the hardness of the resulting optimization problem, the authors propose greedy optimization strategies which they  empirically evaluated on semi-synthetic data. Pricing based on the predicted WTP is also considered in~\citep{tk:epublist:117}.

A quite different optimization problem is formulated in \citep{Hammar:2013:UMC:2507157.2507169}, where the goal is to generate a set of recommendations that maximizes the probability of a purchase.
Therefore, instead of maximizing the revenue based on individual-item profitability considerations, the main short-term goal is to convert the visitor to a buyer. Challenging existing works that solely focus on purchase probabilities, Bodapati \citep{doi:10.1509/jmkr.45.1.77} argues that one should also consider how consumer's would behave if no recommendations would be presented to them. If a certain product will be purchased by a consumer anyway with a certain probability, it might be better not to recommend it, given the limited number of recommendations that can be made. 

Generally, all works discussed so far were mostly evaluated based on some form of simulations based either on synthetic or real-world data. The work of Azaria et al.~\citep{Azaria:2013:MRS:2507157.2507162} is one of the few examples where the consumers' quality perception of and satisfaction with profit-optimized recommendation was assessed in a user study. The participants, who were recruited via Amazon Mechanical Turk, received different types of recommendations and were also queried about their willingness to pay. The results of a real-world field test on the effect of different recommendations, including a profit-oriented one based on the model of \citep{HosanagarICIS2008}, are reported by Panniello et al.~\citep{Panniello201687}. In their study, recommendations were provided to consumers through e-mail newsletters. Both these studies found that a profit-sensitive strategy led to an increased average revenue without a significant loss in terms of the participants' satisfaction.

The study by Panniello et al.~is one of the few works that consider longer-term effects of profit-aware recommendations. Most other works so far in contrast focus on maximizing business value in the short term. Longer-term effects were studied in particular by  Hosanagar et al.~\citep{HosanagarICIS2008} who concluded from their theoretical analyses that optimal recommendations balance profit margins and item relevance. Furthermore, they emphasized on the importance of considering the current reputation of the provider when implementing the strategy.

The expected \emph{customer lifetime value} (CLV) is a well-known instrument from the management and marketing literature. Limited work however exists that tries to connect CLV-related activities like promotions, cross-buying or retention-pricing in combination with recommender systems. Recency, frequency, and monetary (RFM) characteristics of consumers are usually used as a basis for CLV estimates. One possible approach, as discussed in \citep{Liu2005181,Shih2008350}, is to group consumers into segments if their CLV estimates are similar and to incorporate the consumer's segment assignment in the recommendation process. To what extent the consideration of these aspects has an impact on longer-term customer loyalty and the resulting CLV was however not yet been the focus of experimental evaluation.

\subsection{Implementing Value-Aware Recommendation}\label{sec:implementing}
An interesting real-world context for value-aware multistakeholder occurs in multisided platforms where the system earns a variable commission from sales to different providers. For example, a travel web site may earn a commission when a user uses its site to book a hotel room, and these commissions may vary by property. The choice of which hotel rooms to recommend therefore involves a distribution of utility among all three stakeholders. In this section, we discuss a solution developed for this complex scenario. 

A direct approach would be to try to compute and optimize expected conversion rate times expected profit per conversion, a classic economic approach to optimizing revenue.  However, as outlined in~\citep{KrasnodebskiD16} it will generally not give the optimal solution in real world situations. One significant reason is because specific conversion values for recommendations are difficult to predict accurately, much more so than generating a ranked list of recommendations.  As such, the estimated conversion may introduce error into the solution.  In addition, the conversion component is likely to interact with the profit per conversion term.  Because of the complexity of optimizing the equation, the solution is likely to be sub-optimal.

A second approach is to add a separate value-aware function and combine it with the relevancy-based recommendations such that profit is used to distinguish recommendations with similar relevancy.  Conceptually, the designer of the recommendation system can add their knowledge of consumer behavior and profitability of supply to maintain relevancy while trying to maximize profit.  Business criteria such as budget restrictions, competitive offerings, particulars on the profitability of different groups of items can be factored into the value-aware function and how it is combined with the relevancy rankings.  The Kendall-tau metric can be used to optimize the value-aware function by comparing the recommendations based only on relevancy with those combining the value-aware function and minimizing the difference.  As this method is based on heuristics, it needs to be supported with simulations and live testing to ensure the value-aware recommendations maintain a high level of relevancy.

For convenience in this case~\citep{KrasnodebskiD16} found it to best decompose compensation to its base elements, i.e. to the percentage commission and product cost.  These could then be compared and recombined with the other components of the recommendations.  It was also found that balancing the price component was a key element, since higher prices naturally decreased user purchase conversion even while increasing profits.

\subsubsection{Approach}

The approach outlined here is to combine value-awareness into a ranking objective directly as described in \citep{nguyen2017multi}. To do so, the formation of the recommendation list was defined as a multi-objective problem involving a linear combination of the potentially-conflicting objectives of product relevancy, price and the commission percentage earned by the recommender. This was accomplished by defining a novel learning to re-rank optimization problem built on the kernel version of the Kendall tau correlation metric~\citep{jiao2018kendall}.

The concept is essentially an extension of the second approach for value-awareness inclusion, but instead of having a separate value function, Nguyen developed his recommender algorithm to re-rank the initial set of relevancy recommendations with a learning to rank algorithm, so the final ranking was also optimized for the commission, while not deviating significantly from the base ranking. He showed the effectiveness of this approach against a business-rules based model that followed the second approach on a real-world dataset of hotel recommendations from Expedia. It delivered the best performance trade-off for the two objectives under consideration, conversion and margin. This algorithm is detailed in the sections that follow.

\subsubsection{Ranking with sales margin}

Let $\mathbf{X} \in \mathcal{X}$ denote the set of $n$ items with some set of known relevance labels $\mathbf{y} \in \mathcal{Y}$ derived from user interaction Also let $f: \mathcal{X} \times \mathcal{Y} \rightarrow \mathbb{R}$ be the recommender system used at the first stage to produce the vector of scores $\mathbf{u} \in \mathbb{R}^n$ that describe the product's relevance or utility with respect to a consumer search, that is:
\begin{align*}
\mathbf{u} &= \{ \Pr(y_i=1|\mathbf{x}_i) \}_{i=1}^n \\
&= \{ f(\mathbf{x}_i, y_i) \}_{i=1}^n
\end{align*}

These relevance scores are sorted from high to low to deliver a relevance ranking of product recommendations. We will use the notation $r(u_i)$ to indicate the rank of item $i$ in such a list by virtue of having the score $u_i$. To produce such scores, learning-to-rank recommender systems typically follow a pairwise approach in which a ranking metric like the Normalized Discount Cumulative Gain (NDCG) is optimized over pairs of items so that relevant items are ranked in top positions ~\citep{burges2010ranknet,burges2005learning,burges2011learning}. In particular, LambdaRank~\citep{burges2010ranknet} is one of the most effective approaches~\citep{donmez2009local} due to its ability to minimize a surrogate loss $\mathcal{L}_r$ of the NDCG metric (equivalent to maximizing this metric):
\begin{align}
\label{f:Lr}
\min_f \mathcal{L}_r(\mathbf{y} | \mathbf{X}) = \sum_{y_i \geq y_j} \log(1 + \exp( -\theta (u_i - u_j) )) |\Delta_{ij}^ \text{NDCG}|
\end{align}
where $u_i = f(\mathbf{x}_i, y_i)$ and $\Delta_{ij}^\text{NDCG} = (2^{y_i} - 2^{y_j}) (\log(1+r(u_i)) - \log(1 +r(u_j)))$ is the cost in terms of NDCG of exchanging item $i$ with item $j$ from position $r(u_i)$ to position $r(u_j)$.

The first step in adding value-awareness to $f$ is to outline the commission paid to the recommender as a portion of sales (margin, $m$), which is equal to the selling price $p$ of the product minus its cost $c$, i.e. $m = p - c,$ and in the expectation is equal to:
\begin{align}
\label{f:E}
\E[m|\mathcal{X},\mathcal{Y}] &= \E[ \Pr(\mathcal{Y}=1|\mathcal{X}) \times m ] \nonumber \\ 
&= \E[ f(\mathcal{X}, \mathcal{Y}) \times m ]
\end{align}

\noindent To maximize its commission the recommender has then to solve the following objective function $\mathcal{L}$ given by maximum log-likelihood of (\ref{f:E}):
\begin{align}
\label{f:L}
\max_{\alpha,\beta} \mathcal{L}(\mathbf{m}|\mathbf{u}) &= \sum_{i=1}^n \log(u_i) + \alpha \log(p_i) + \beta \log( m_i / p_i) 
\end{align}
where $\alpha$ and $\beta$ are tuning parameters (originally set to one) that gives the importance of the log margin component which are separated into price and margin percent; i.e. $\log(m) = \log(m/p) + \log(p)$. 

As initially noted, there is the potential for interactions between these factors and the recommendation task is a multi-objective optimization problem fitting the $\langle C^-_p, P^-_n, S_t \rangle$ description. The objectives include relevance as defined by the score $u$, the supplier's interest as defined by price $p$ and the intermediary's percent commission $m/p$. The core problem is to solve for the intrinsic relation between the factors, which are contradictory: consumers want the lowest price, sellers want the highest price and lowest commission, and the platform wants the highest commission consistent with consumer purchase). Note that this is not a simple aggregation $S_a$ of supplier margins because the commissions may differ among them.

The problem then becomes how to optimize the weight $\beta$ of the recommender with respect to the input attribute $\mathbf{X}$, the relevance scores $\mathbf{u}$ and the suppliers' margins $\mathbf{m}$, that is:
\begin{align}
\label{f:L3}
\max_\beta \mathcal{L}(\mathbf{m}|\mathbf{u},\mathbf{X}) &= \sum_{i=1}^n \log(u_i) + \alpha \log(p_i) + \beta(\mathbf{x}_i,m_i) \log( m_i / p_i)
\end{align} 
where $\alpha$ was treated as a fixed hyper-parameter. 

\subsubsection{Learning to re-rank}

The problem in Equation~\ref{f:L3} was framed as a learning-to-re-rank problem such that the new ranking order provided by this formula stays as close as possible to the original ranking order $\mathbf{u}$ while maximizing the margins $\mathbf{m}$. This was done using the Kendall tau correlation metric, which operates on the relative pairwise order of the items to produce a new ranking of items which maximizes the margin while minimizing the distance between the new and old rankings.

$\mathbf{u}' \in \mathbb{R}^n$ is the set of new scores derived from equation (\ref{f:L3}), i.e. 
\[ \forall u \in \mathbf{u} \text{, } u' = u + \alpha \log(p) + \beta(\mathbf{x}, m) \log(m/p) \] 
    
From a ranking perspective, $\mathbf{u}'$ is one permutation of $\mathbf{u}$ in the $\binom{n}{2} = n (n-1) / 2$ permutation space of $n$ items, and to measure the distance between the two score vectors in this space the Kendall tau correlation measure~\citep{kendall1938new} can be used. It counts the number $n_c$ of concordant pairs versus the number $n_d$ of discordant pairs between the two vectors can be used. Thus we can compute:
\begin{align*}
n_c(\mathbf{u}, \mathbf{u}') =& \sum_{i<j} \I(r(u_i) > r(u_j)) \I(r(u'_i) > r(u'_j)) \\
& + \I(r(u_i) < r(u_j)) \I(r(u'_i) < r(u'_j)) 
\end{align*}
\begin{align*}
n_d(\mathbf{u}, \mathbf{u}') =& \sum_{i<j} \I(r(u_i) > r(u_j)) \I(r(u'_i) < r(u'_j)) \\
& + \I(r(u_i) < r(u_j)) \I(r(u'_i) > r(u'_j)) 
\end{align*}
The Kendall tau correlation measure can then be defined as:
\begin{align}
\label{f:K}
K(\mathbf{u}, \mathbf{u}') &= \frac{n_c(\mathbf{u}, \mathbf{u}') - n_d(\mathbf{u}, \mathbf{u}')}{n (n-1) / 2} 
\end{align}
which has been shown to be a positive-definite kernel~\citep{jiao2018kendall}. 

With such a kernel, we can apply the kernel trick ~\citep{hofmann2008kernel} to optimize over this metric. More precisely, following ~\citep{jiao2018kendall}, we define the mapping function $\phi: \mathbb{R}^n \rightarrow \mathbb{R}^{n (n-1) / 2} $, as well as its smooth version $\phi'$, which is given by transforming the indicator function $\I$ of the function $\phi$ to its sigmoid counterpart $\sigma(x) = \frac{1}{1+\exp(- \theta x )}$:
\begin{align*}
\phi(\mathbf{u}) &= \left( \frac{1}{\sqrt{n (n-1) / 2}} ( \I(r(u_i) > r(u_j)) - \I(r(u_i) < r(u_j)) ) \right)_{i<j} \\
\phi'(\mathbf{u}) &= \left( \frac{1}{\sqrt{n (n-1) / 2}} ( \sigma(u_i - u_j) - \sigma(u_j - u_i) ) \right)_{i<j} 
\end{align*}
from which the kernelized version of Kendall tau can be defined as:
\begin{align}
\hat K(\mathbf{u}, \mathbf{u}') = \phi(\mathbf{u})^{\T} \phi'(\mathbf{u}')
\end{align}
which makes use of the smooth $\phi'$ on the new score $\mathbf{u}'$ since learning will occur on this side of the kernel. 

The task is two-fold for the optimization problem: learn a new ranking order $\mathbf{u}'$ of the items, described by features $\mathbf{X}$, which maximizes the NDCG of commissions $\mathbf{m}$. This can be expressed as a standard learning-to-rank problem similar to equation (\ref{f:Lr}), but defined here as $\mathcal{L}_r(\mathbf{m} | \mathbf{X})$ and optimized for the new scores $\mathbf{u}'$. At the same time, the distance of the new ranking $\mathbf{u}'$ to the original ranking $\mathbf{u}$ is to be minimized, which can be written using the Kendall tau metric $K(\mathbf{u}, \mathbf{u}')$. The full optimization problem can be then written as: 
\begin{align}
\label{best_b}
&\min_{\beta} \mathcal{L}(\mathbf{m} | \mathbf{u}, \mathbf{X} ) = \mathcal{L}_r(\mathbf{m} | \mathbf{X}) + \gamma (1-\hat K(\mathbf{u}, \mathbf{u'})) 
\end{align} 

In problem (\ref{best_b}), the kernelized Kendall tau metric $\hat K(\mathbf{u}, \mathbf{u}')$ plays the role of a similarity-based regularizer with the original ranking order $\mathbf{u}$ being the reference point to the new ranking order $\mathbf{u}'$. Also, the hyper-parameter $\gamma$ gives the balance between the two terms of the optimization; with a high $\gamma$, the new ranking order $\mathbf{u}'$ will not diverge too much from the original one, but neither will the commission, while with a low $\gamma$ items with a high commission will be more likely to be pushed up, increasing the chance of profit, unless the items have a low relevance.
 
Overall, the problem (\ref{best_b}) is a reformulation of the multi-objective problem (\ref{f:L3}) in the learning-to-rank framework; it finds exactly the best weight $\beta$ that maximizes the commission with the constraint of maintaining the importance of relevance $\mathbf{u}$. 
Therefore by solving problem (\ref{best_b}), (\ref{f:L3}) is also solved. 

\subsubsection{Experimental Evaluation}

The learning-to-re-rank (LRR) approach was evaluated on an Expedia dataset built from world-wide hotel searches collected during 2016, where the underlining recommendation technique was matrix factorization ~\citep{Abernethy2006}. The baseline value-aware model was based on one developed using the second value-awareness method and had been optimized through extensive A/B testing.
 ~\citep{krasnodebski2016considering}.
Both of these were compared to the relevance-only recommendations, which had no value-awareness (based purely on $u_i$),

The results are provided in Table \ref{tab:ndcg}.  NDCG@10 on customer clicks + bookings was used to measure the relevance of the recommendations and the achieved commission \% for the value-awareness. We see that both value-aware approaches delivered similar performance for the commission, +16.7\%, but the learning to re-rank approach has a lower loss in relevance , with a decrease of -5.9\% versus -8.7\% in NDCG.


\begin{table}
\centering
\caption{Relative Improvement of NDCG@10 on Expedia Hotel Searches versus Relevance-only recommendations using Learning to Re-Rank}
\label{tab:ndcg}
\begin{tabular}{|c|c|c|}
\hline
& Clicks+Bookings & Margin \\ \hline
Baseline & -8.7\% & +16.7\% \\ \hline
LRR & -5.9\% & +16.7\% \\ \hline
\end{tabular}
\end{table}

\subsection{Summary}
Our brief review and example shows that there is a history of both applied and theoretical work in recommender systems in which the roles and goals of the different stakeholders are considered in developing of price- and profit-aware recommender systems. The reviewed studies in general suggest that profit-aware recommendation strategies can lead to a substantially higher business values for the provider, at least in the short term, without a considerable loss in recommendation quality and trust. The proposed models however differ in their complexity. While some studies are based on comparably simple, static adoption probabilities, others consider more comprehensive models with time-varying adoption probabilities and limited consumer budgets.  Furthermore, some works focus on promoting the most profitable products through recommendations, others aim to maximize the profit by trying to maximize the conversion rate or by stimulating purchases which would not have happened without the recommender system. The example application demonstrates a real-world context in which multiple stakeholder considerations are relevant and can be combined successfully into a single optimization model.

\section{Example: Fairness-aware Recommendation}
    \subsection{Fairness-aware Recommendation}
\label{sec:far}


In this section, we discuss fairness in a recommendation context.
In this context, the term \emph{fairness} has two meanings.
The one is fairness in recommendation process, and the other is fairness of resource allocation.
The former means that specific information does not influences a recommendation outcome, and the information is ignored in the recommendation process.
The latter means that resources to be recommended are fairly allocated to stakeholders.

\subsubsection{Fairness in Recommendation Process}
\label{sec:far:process}

In this section, the term \emph{fairness} means that the specified information is not exploited when generating a recommendation outcome.
For example, when recommending jobs, if information about job-applicant's gender does not influence a recommendation outcome, the outcome is considered fair in a sense that the socially sensitive information is not abused to determine what jobs are recommended to the applicant.
Hereafter, we represent such information to be ignored by a variable, which is called a sensitive feature.
Such a type of fairness is firstly discussed by \cite{tk:kdd:08:03}

There are several definitions of formal fairness in this sense, which are distinguished by the aspect of a recommendation outcome influenced by a sensitive feature.
We first introduce these formal definitions of fairness.
Then, we show applications of this concept for solving real problems.
We finally introduce algorithms to enhance these fairness in a recommendation context.

\paragraph{Formal Definitions}
\label{sec:far:process:definition}

After defining notations, we enumerate some definitions.
Consider an event in which all the information required to make a recommendation, such as the specifications of a user and item and all features related to them, is provided and a recommendation result is inferred from this information.
This event is represented by a triplet of three random variables: $Y$, $S$, and $X$.
$Y$ represents a feedback of a user, which is typically a rating value or an indicator of whether or not a specified item is relevant.
$S$ stands for a sensitive feature.

Finally, $X$ represents all ordinary features (or features) related to this event other than those represented by $Y$ and $S$.
A task of recommendation needs to predict a feedback for an item for which an active user has not yet give a feedback.
We call such a predicted feedback by a recommendation outcome, and it is denoted by $\hat{Y}$.

We next move on to notions of fairness in a context of machine learning or data mining~\cite{tk:misc:179}.
Because so many indexes to quantify the degree of fairness have been developed~\cite{tk:dmkd:17:01}, it is unrealistic to enumerate these notions here.
Instead of these indexes, we introduce the mathematical conditions that the indexes is designed to represent.
For example, the difference and the ratio between the probabilities conditioned by distinct sensitive values, $\Pr[\hat{Y}{=}1 | S{=}0] - \Pr[\hat{Y}{=}1 | S{=}1] \rightarrow 0$ and $\Pr[\hat{Y}{=}1 | S{=}0] / \Pr[\hat{Y}{=}1 | S{=}1] \rightarrow 1$, are proposed as indexes for measuring fairness.
If both of these indexes take their corresponding ideal value, a mathematical condition of the statistical independence between, $\hat{Y}$ and $S$, $\hat{Y} \indep S$, is satisfied.
Because multiple indexes can be often aggregated into a single condition, we can discuss formal notions of fairness more simply.
Note that $A \indep B$, introduced by Dawid, denotes the (unconditional) independence between variables $A$ and $B$, and $A \indep B \mid C$ denotes the conditional independence between $A$ and $B$ given $C$.

Now, we enumerate three major formal conditions of fairness.
The first one is conditional independence between $\hat{Y}$ and $S$ given $X$, $\hat{Y} \indep S \mid X$.
This condition is related with notions called by \emph{direct fairness} or \emph{disparate treatment} in a literature of fairness-aware machine learning.
The recommendation is made by inferring the value of $Y$ given the values of $S$ and $X$ based on a probabilistic recommendation model, $\Pr[\hat{Y} | S, X]$.
It might appear that the model could be made independent by simply removing $S$, but this is not the case.
By removing the sensitive information, the model satisfies the condition $\Pr[\hat{Y} | S, X] = \Pr[\hat{Y} | X]$.
Using this equation, the probability distribution over $(Y, S, X)$ leads conditional independence, $\hat{Y} \indep S \mid X$.
\begin{align*}
\Pr[\hat{Y}, S, X] = \Pr[\hat{Y} | S, X] \Pr[S | X] \Pr[X]
= \Pr[\hat{Y} | X] \Pr[S | X] \Pr[X]
.
\end{align*}
Under this condition, if there are features in $X$ that are not independent of $S$, the outcomes will be influenced by $S$ through the dependent features.
For example, even though no information about job-applicants' race was explicitly exploited, their home address contains information about their race if distinct races live separately apart.
Note that such an indirect influence is called a red-lining effect~\cite{tk:dmkd:10:01}.

\begin{figure}
\centering
\subcaptionbox{$\hat{Y} \indep S$\label{tk:fig:indep-criteria:a}}%
{\includegraphics[width=0.35\linewidth]{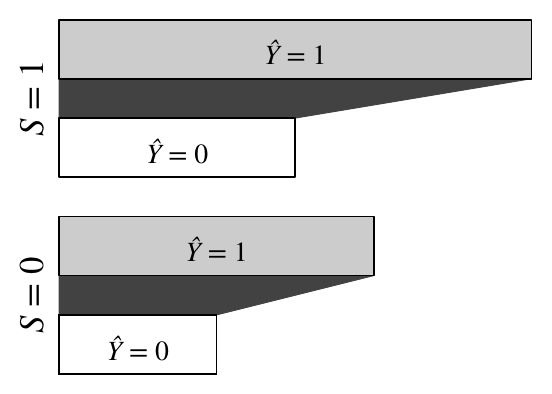}}%
\hspace{0.1\linewidth}%
\subcaptionbox{$\hat{Y} \indep S \mid Y$\label{tk:fig:indep-criteria:b}}%
{\includegraphics[width=0.35\linewidth]{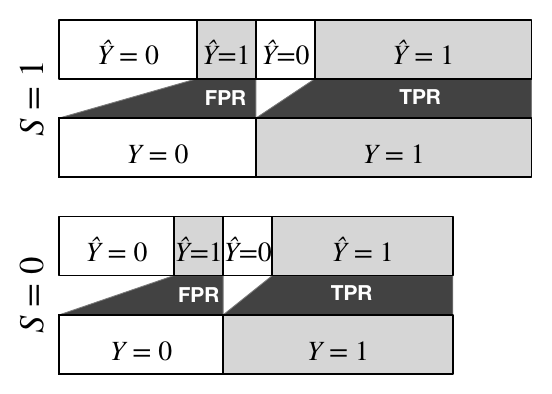}}%
\caption{Distributions of the predicted ratings for each sensitive value}
\label{tk:fig:indep-criteria}
\end{figure}

The second condition is unconditional independence between $\hat{Y}$ and $S$, $\hat{Y} \indep S$.
This is related to notions known as statistical parity or independence.
Unlike the first condition, the indirect influence of a sensitive feature, as well as its direct influence, can be removed by satisfying this independence condition.
From a viewpoint of information theory, this condition equivalent to the condition that mutual information between $\hat{Y}$ and $S$ is zero.
This means that we know nothing about $\hat{Y}$ even if we obtain information about $S$.
Figure~\ref{tk:fig:indep-criteria:a} illustrates the case both an outcome and a sensitive feature are binary variables.
Ratios of $\hat{Y}{=}0$ to $\hat{Y}{=}1$ (depicted by black areas) in the upper and lower figures should be matched to satisfy this type of fairness.

The final condition is conditional independence between $\hat{Y}$ and $S$ given $Y$, $\hat{Y} \indep S \mid Y$.
This is related to notions known as equalized odds or sufficiency~\cite{tk:nips:17:02}.
The above two conditions assumes that feedbacks from users are potentially unfair.
For example, minority individuals have been more poorly rated than their actual ability in job recommendation.
On the other hand, feedbacks are assumed to be fair and reliable in this condition~\cite{tk:www:17:01}.
This condition indicates that errors of predictions, $\hat{Y}$, against observations, $Y$, are not influenced by a sensitive information, $S$.
Even if feedbacks are fair, prediction errors can be influenced by a sensitive feature due to inductive bias of a prediction algorithm.
Such influence can be removed by satisfying this condition.
In Figure~\ref{tk:fig:indep-criteria:b}, true positive ratio (TPR) and false positive ratio (FPR) in the upper and lower figures should be matched to satisfy this type of fairness.

We finally comment on several conditions representing fairness.
Context-specific independence, $\hat{Y} \indep S \mid X{=}x$, is fairness satisfied in a specific context so that $X$ takes $x$.
This is a weaker condition of $\hat{Y} \indep S \mid X$, and is used in a task of discovering unfair decisions in a database~\cite{tk:kdd:08:03}.
A weaker condition of $\hat{Y} \indep S$ is that $\hat{Y}$ and $S$ are uncorrelated~\cite{tk:ec:049}.
Envy-free, which is notion of fair resource allocation in a game theory, is applied to fairness-aware data mining~\cite{tk:nips:18:02}.

\paragraph{Applications}
\label{sec:far:process:application}

We here consider three types applications for which the above fairness conditions would be useful.
From a viewpoint of multistakeholder recommendation, the first and third applications fair for consumers, while the second application is fair for providers.

\noindent[Adherence to Laws and Regulations]
Recommendation services must be managed while adhering to laws and regulations.
We will consider the example of a suspicious advertisement placement based on keyword-matching~\cite{tk:macm:13:01}.
In this case, users whose names are more popular among individuals of African descent than European descent were more frequently shown advertisements implying arrest records.
According to an investigation, however, no deliberate manipulation was responsible; rather, the bias arose simply as a side-effect of algorithms to optimize the click-through rate.
Because similar algorithms of Web content optimization are used for online recommendations, such as for online news recommendations, similar discriminative recommendations can be provided in these contexts.
For example, fairness-aware recommendation would be helpful for matching an employer and a job applicant based not on gender or race, but on other factors, such as the applicant's skill level at the tasks required for the job.

Fairness-aware recommendation is also helpful for avoiding the use of information that is restricted by law or regulation.
For example, privacy policies prohibit the use of certain types of information for the purpose of making recommendations.
In such cases, by treating the prohibited information as a sensitive feature, the information can be successfully excluded from the prediction process of recommendation outcomes.

\noindent[Fair Treatment of Content Providers]
Fairness-aware recommendation can be used to ensure the fair treatment of content providers or product suppliers.
The Federal Trade Commission has been investigating Google to determine whether the search engine ranks its own services higher than those of competitors~\cite{tk:misc:139}.
The removal of deliberate manipulation is currently considered to ensure the fair treatment of content providers.
However, algorithms that can explicitly exclude information whether or not content providers are competitors would be helpful for dismissing the competitors' doubts that their services may be unfairly underrated.

Though this case is about information retrieval, the treatment of content providers in the course of generating recommendations can also be problematic.
Consider the example of an online retail store that directly sells items in addition to renting a portion of its Web sites to tenants.
On the retail Web site, if directly sold items are overrated in comparison to items sold by tenants, then the trade conducted between the site owner and the tenants is considered unfair.
To carry on a fair trade, the information on whether an item is sold by the owner or the tenants should be ignored.
Enhancement of fairness  would be helpful for this purpose.

\noindent[Exclusion of Unwanted Information]
Users may want recommenders to exclude the influence of specific information.
We give several examples.
Enhancing fairness is useful for correcting a popularity bias, which is the tendency for popular items to be recommended more frequently~\cite{tk:kdd:08:04}.
If users are already familiar with the popular items and are seeking minor and long-tail items that are novel to them, this popularity bias will be unfavorable to them.
In this case, the users can specify the volume of consumption of items as a sensitive feature, and the algorithm will provide recommendations that are free from information about the popularity of items.


When users explicitly wish to ignore specific information, such information can be excluded by enhancing fairness.
Pariser recently introduced the concept of the filter bubble problem, which is the concern that personalization technologies narrow and bias the topics of interest provided to technology consumers, who do not notice this phenomenon~\cite{tk:misc:005}.
If a user of a social network service wishes to converse with people having a wide variety of political opinions, a friend recommendation that is not influenced by the friends' political conviction will provide an opportunity to meet people with a wide range of views.

\paragraph{Algorithms}
\label{sec:far:process:algorithm}

We here introduce recommendation algorithms that enhance three types of fairness.
First, we discuss $\hat{Y} \indep S \mid X$.
Basically, this condition can be satisfied simply by removing a sensitive information from a prediction model.
We here introduce a model for predicting how much movie's popularity, the number of views, increases if the movie is awarded~\cite{tk:lncs:17:03}.
In this case, a sensitive feature indicates whether or not a movie is awarded.
The increase of movie's popularity is assumed to be the difference between the popularity of awarded and non-awarded movies.
If a sensitive feature is sufficiently independent from movie's other features, this assumption would be valid, and this model becomes free from a red-lining effect.

We next move on to the second condition, $\hat{Y} \indep S$.
We have been developed two approaches to enhance this type of fairness.
The first regularization approach is originally proposed for classification~\cite{kamishima2012fairness}.
This approach adopts a regularizer imposing a constraint of independence while training a recommendation model~\cite{tk:epublist:126}.
A basic form of an objective function is
\[
\loss(Y, \hat{Y}) - \eta\,\mathrm{indep}(\hat{Y}, S) + \lambda\,\mathrm{reg}
,
\]
where $\loss(\cdot)$ is an empirical loss and $\mathrm{reg}$ is a regularizer to avoid overfitting.
$\mathrm{indep}(\cdot)$ is an independence term, and it takes the larger value if an outcome is fairer.
An independence parameter, $\eta$, controls the balance between prediction accuracy and fairness.
We define $\calD^{(s)}$ as a subset consisting of all training data whose sensitive feature takes $s$.
Our first efficient independence term is mean matching, which is designed to match means of two datasets, $\calD^{(0)}$ and $\calD^{(1)}$, 
\[
- \prn{\mean(\calD^{(0)}) - \mean(\calD^{(1)})}^{2}
.
\]
However, this term ignores the second moment of rating distributions.
We developed two independence terms that can take the second moments into account~\cite{tk:epublist:197}.
The one term is distribution matching with Bhattacharyya distance,
\[
- \prn{- \ln \int \sqrt{\Pr[\hat{Y} | S{=}0] \Pr[\hat{Y} | S{=}1]}}
,
\]
which measures the distance between two distributions, $\Pr[\hat{Y} | S{=}0]$ and $\Pr[\hat{Y} | S{=}1]$, in Bhattacharyya distance.
The other term is mutual information with normal distributions,
\[
\textstyle
-\prn{\ent(\hat{Y}) - \sum_{s} \Pr[S{=}s] \ent(\hat{Y} | S{=}s)}
,
\]
where $\ent(\cdot)$ is a differential entropy function for a normal distributions.

\begin{figure}
\centering
\subcaptionbox{standard recommendation\label{tk:fig:score-dist:a}}%
{\includegraphics[width=0.4\linewidth]{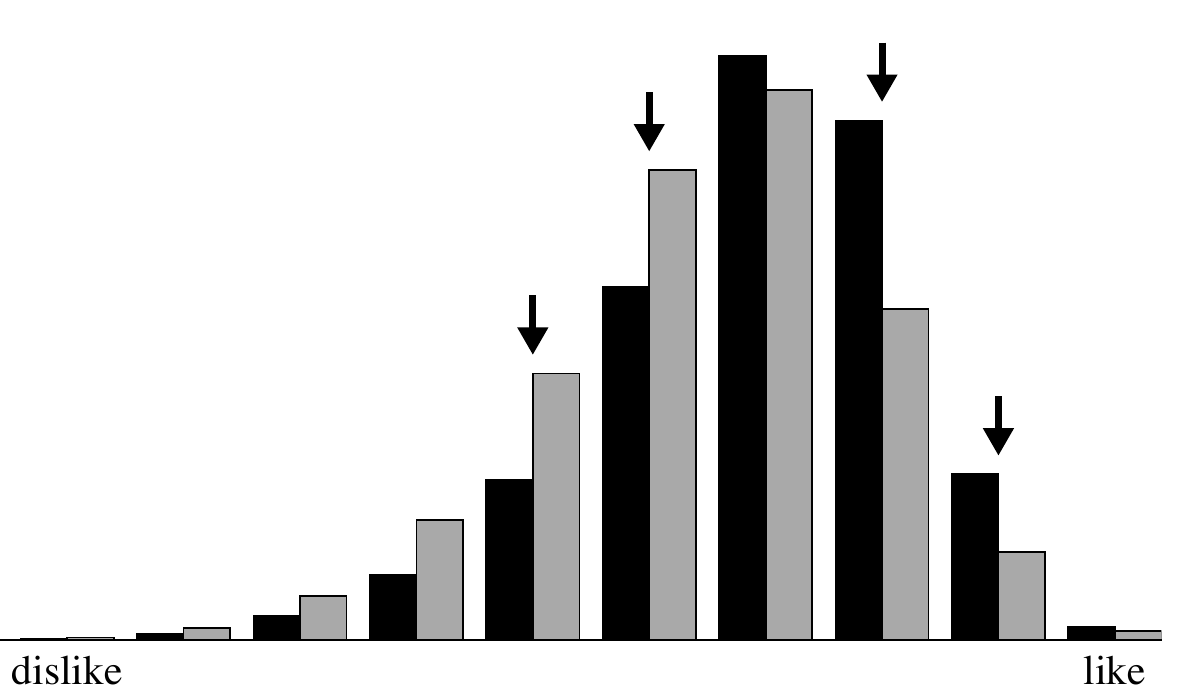}}%
\hspace{0.1\linewidth}%
\subcaptionbox{fairness-aware recommendation\label{tk:fig:score-dist:b}}%
{\includegraphics[width=0.4\linewidth]{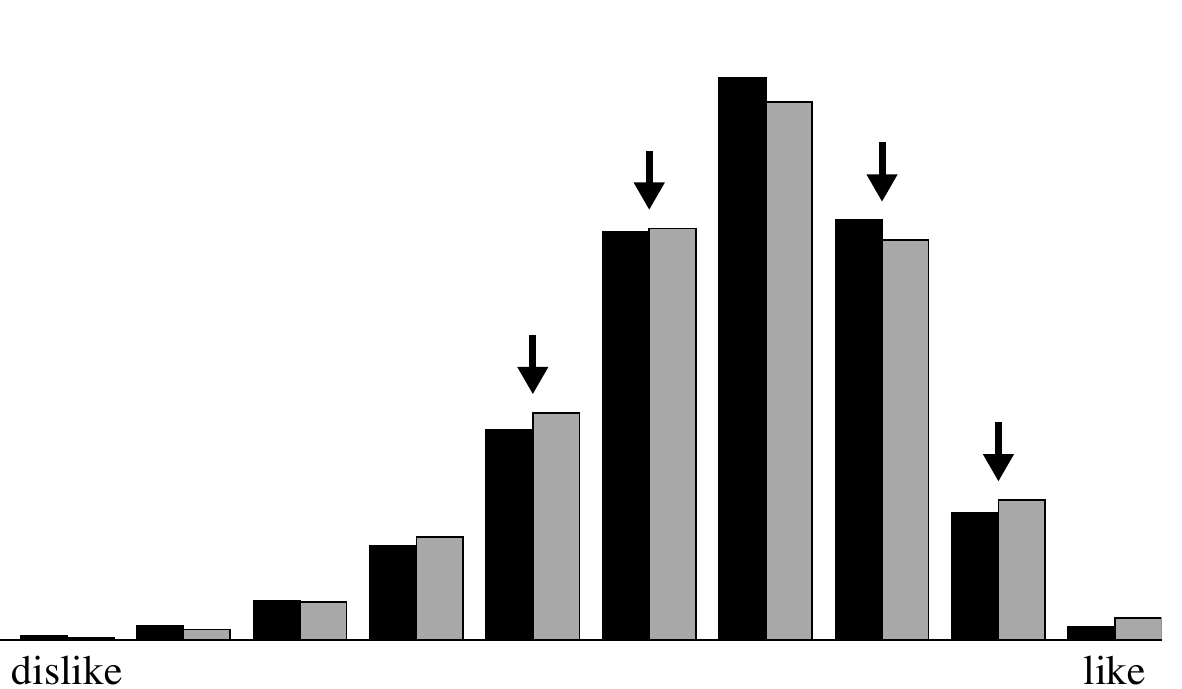}}%
\caption{Distributions of the predicted ratings for each sensitive value}
\label{tk:fig:score-dist}
\end{figure}

We here demonstrate the effect of our fairness enhancement.
Figure~\ref{tk:fig:score-dist} shows the distributions of predicted ratings for movies.
In this example, a sensitive feature represents whether or not movies are released before 1990.
Black and gray bars show the distributions of ratings for older and newer movies, respectively.
In Figure~\ref{tk:fig:score-dist:a}, ratings are predicted by a standard algorithm, and older movies are highly rated (note the large gaps between the two bars indicated by arrowheads).
When fairness is enhanced as in Figure~\ref{tk:fig:score-dist:b}, the distributions of ratings for older and newer movies become much closer (the large gaps are lessened); that is to say, the predicted ratings are less affected by movies' release-year.

The second approach of enhancing independence between $\hat{Y}$ and $S$ is a model-based approach~\cite{tk:epublist:183}.
In this approach, a probabilistic generative model for collaborative filtering is extented~\cite{tk:ijcai:99:01},
\[
\textstyle
\Pr[X, Y] = \sum_{Z} \Pr[X | Z] \Pr[Z | Y] \Pr[Y]
.
\]
$Z$ is a latent variable, which is introduced to simplify the model.
To this model, we add a sensitive feature so that it is independent from an outcome variable,
\[
\textstyle
\Pr[X, Y, S] = \sum_{Z} \Pr[X | Z, S] \Pr[Z | Y, S] \Pr[Y]\Pr[S] = \Pr[X | Y, S] \Pr[Y] \Pr[S]
.
\]
In our experiment, this approach was inferior to the above regularization approach in terms of fairness, because the assumption of this model is unrealistic.
Independence between $\hat{Y}$ and $S$ is satisfied if a predicted outcome is assumed to be probabilistically following a distribution represented by this model.
However, an actual outcome is an expectation of $\Pr[\hat{Y} | X, S]$, which is deterministically~generated~\cite{tk:epublist:190}.
This discrepancy between the model's assumption and an actual condition worsens fairness.

The final fairness is $\hat{Y} \indep S \mid Y$ is considered in~\cite{tk:nips:18:01}.
The proposed approach is similar to the above regularization approach, but an independence term is replaced.
Absolute unfairness is a measure of unfairness,
\[
\mean_\mathrm{user} \Big|
\big| \mean_\mathrm{item0}[\hat{Y}_\mathrm{user}] - \mean_\mathrm{item1}[\hat{Y}_\mathrm{user}] \big| -
\big| \mean_\mathrm{item0}[Y_\mathrm{user}] - \mean_\mathrm{item1}[Y_\mathrm{user}] \big| 
\Big|
.
\]
Item sets are divided into two sensitive groups each of which shares the same sensitive value, and these are indicated by $\mathrm{item0}$ and $\mathrm{item1}$.
The first inner absolute is the difference between means of predicted outcomes for a specific user over two item groups.
The second inner absolute is the difference between means of observed outcomes.
This absolute unfairness measures the dissimilarity between these two differences.



\section{Methodological Issues}
    At this point in the development of multistakeholder recommendation research, there is a diversity of  methodological approaches and little agreement on basic questions of evaluation. In part, this is a reflection of the diversity of problems that fall under the multistakeholder umbrella. It should be noted that evaluation can be both formative and summative in recommender systems research: we evaluate in order to improve systems, choosing ``the best'' among different algorithmic variants and parameter settings. We also evaluate in order to show the impact (or more often the predicted impact) of our system on its stakeholders. 

A key difficulty is the limited availability of real-world data with multistakeholder characteristics. The reason for this becomes clear if we consider the experiments shown in Section \ref{sec:implementing}. The data that makes these experiments possible is highly business-critical, including such data as the margins associated with each provider and the variable commissions negotiated by the platform. It is unlikely that an organization would be able to share this kind of data with outside researchers. Some researchers have obtained proprietary data for multistakeholder research, but progress in the field requires replicable experiments that proprietary data does not support. Areas of multistakeholder research that involve public, rather than private, benefit may offer advantages in terms of the availability of data: see, for example, the data sets available from the crowd-funded educational charity site DonorsChoose.org\footnote{https://data.donorschoose.org/explore-our-impact/}.

\subsection{Simulation}

In the absence of real-world data with associated valuations, researchers have typically turned to simulations as in Section~\ref{sec:balanced} above. Simulated or inferred provider data is useful for transforming publicly-available recommendation data sets in standard user, item, rating format into ones that can be used for multistakeholder experimentation. The experiments in \citep{surer2018multistakeholder} provide an example of this methodology: each item in the MovieLens 1M data set was assigned to a random provider, and the distribution of utilities calculated. To capture different market conditions, the experimenters use two different probability distributions: normal and power-law. There is no accepted standard for producing such simulations and what are reasonable assumptions regarding the distribution of provider utilities or the formulation of system utilities, except in special cases.  

Researchers have also used objective aspects of data sets to infer proxy attributes for multistakeholder evaluation. In \citep{soappaper} where the first organization listed in the production credits for each movie was treated as the provider -- a significant over-simplification of what is a very complex system of revenue distribution in the movie industry. In other work, global metrics such as network centrality \citep{ValuePick2010} have been used to represent system utility for the purposes of multistakeholder evaluation. \citep{burke2018synthetic} demonstrated an alternate approach to generate synthetic attribute data based on behavioral characteristics that can be used to evaluate system-level fairness properties.

More sophisticated treatments of profitability and recommendation are to be found in the management and e-commerce literature, some using public data as seen in  \citep{Oestreicher-Singer:2012:VHD:2398302.2398303,ChenWuYoon:04a,DBLP:conf/icis/AdamopoulosT15}, but these techniques and associated data sets have not yet achieved wide usage in the recommender system community. 

\subsection{Models of utility}
A multistakeholder framework inherently involves the comparison of outcomes across different groups of individuals that receive different kinds of benefits from the system. In economic terms, this entails utility calculation and comparison. As with data, different researchers have made different assumptions about what types of utilities accrue from a recommender system and how they are to be measured. A standard assumption is that the output of a recommendation algorithm can be treated as approximations of user utility. Yet, research has confirmed that users prefer diverse recommendation lists \citep{pu2011user}, a factor in tension with accuracy-based estimates of user utility. 

Most of the examples discussed above, focus solely on the short-term perspective. More research is therefore required to understand the potential positive and negative long-term effects of profit-aware recommendation and other strategies that are not strictly user-focused. Future models could also consider the price sensitivity and willingness-to-pay of individual consumers in the recommendation process. 

\subsection{Off-line experiment design}
A standard off-line experimental design in recommender systems is the creation of multiple folds of training and test data from a data set using random fixed-sized partitioning of user profiles. The benefit of this approach is that each partition contains a fixed proportion of each user's profile, guaranteeing a minimum profile size for recommendation generation. This makes sense when user outcomes are the highest priority, as it ensures that an evaluation data point can be produced for every user in every fold. All of recommendations for a test fold are produced, in some sense, simultaneously, as a set of recommendation lists or rating predictions to which evaluation metrics can be applied. 

This experimental framework makes a bit less sense in a multistakeholder context, and this is where the essential asymmetry of the stakeholders comes into play. Providers are, in a key sense, passive -- they have to wait until users arrive at the system in order to have an opportunity to be recommended. The randomized cross-fold methodology measures what the system can do for each user, given a portion of their profile data, the potential benefit to be realized if the user visits and a recommendation list is generated. Evaluating the provider side under the same conditions, while a commonly-used methodology, lacks a similar justification.

A more realistic methodology from the provider's point of view is a temporal one, that takes into account the history of the system up to a certain time point and examines how provider utilities are realized in subsequent time periods. See \citep{Campos2014} for a comprehensive discussion of time-aware recommender systems evaluation. However, time-aware methods have their own difficulties, forcing the system to cope with cold-start issues possibly outside of the scope of a given project's aims.

\subsection{User studies}
User studies are another instrument available to researchers that has not been extensively applied to multistakeholder recommender systems. As usual for such studies, the development of reliable experimental designs is challenging as the participants' decision situation typically remains artificial. Furthermore, as in the study by Azaria et al.~discussed above \citep{Azaria:2013:MRS:2507157.2507162}, familiarity biases might exist -- in their study participants were willing to pay more for movies that they already knew -- which have been observed for other types user studies in the recommender systems domain \citep{JannachLercheEtAl2015a}. Ultimately, more field tests -- even though they are typically tied to a specific domain and specific business model -- are needed that give us more insights into the effects of multistakeholder recommendations in the real world.

\subsection{Evaluation Metrics}
The building block of the multistakeholder evaluation is to first measure the utility each of the stakeholders gets within a recommendation platform. Common evaluation metrics such as RMSE, precision, NDCG, diversity, etc. are all different ways to evaluate the performance of a recommender system from the user's perspective. As noted above, these measures are implicitly a form of system utility measure as well: system designers optimize for such measures under the assumptions that (1) higher evaluation metrics correspond to higher user satisfaction and (2) higher user satisfaction contributes to higher system utility through customer retention, trust in the recommendation provided, etc. However, the formulation of multistakeholder recommendation makes it possible to characterize and evaluate system utility explicitly. 

Typically, evaluation metrics are averaged over all users to generate a point score indicating the central tendency over all users. However, it is also the case that in a multistakeholder environment additional aspects of the utility distribution may be of interest. For example, in an e-commerce context, providers who receive low utility may leave the eco-system, suggesting that the variance of provider utility may be important as well as the mean. One suggested practice would be to report the first three moments of the distribution of utilities for each shareholder -- the mean, variance, and skewness -- rather than just the mean value when reporting on multistakeholder evaluations. 

\subsubsection{Provider metrics}
When evaluating the utility of a recommender system for a particular provider, we may take several different stances. One views the recommender as a way to garner user attention. In this case, the relevance of an item to a user may be a secondary consideration. Another perspective views the recommender as a source of potential leads. In this view, recommending an item to uninterested users is of little benefit to the provider.  In the first situation, simply counting (with or without a rank-based discount) the number of times a provider's products appears in recommendation lists would be sufficient. In the second situation, the metric should count only those recommendations that were considered ``hits'', those that appear positively rated in the corresponding test data.

Another provider consideration may be the reach of its recommended products across the user population. A metric could count the number of unique users to whom the provider's items are recommended. As noted in the discussion above the distinction between the $P^-$ and $P^+$ multistakeholder configurations lies precisely in the providers' ability to target specific audiences for their items. In a $P^+$ system, it would make sense to consider reach relative to the target population. For example, in an online dating application where the user can specify desired properties in a match, an evaluation metric might be the fraction out of the target audience receiving the recommendation. 

Finally, where the consideration is the accuracy of system's predictions, we can create a provider-specific summary statistic of a measure like RMSE. \citep{ekstrand2018exploring} uses this method to examine differences in error when recommending books by male and female authors. Since the statistic by itself is not that useful for a single provider, a better metric would indicate the provider's position relative to other providers in the overall distribution

Table~\ref{tab:provider} shows example metrics for each of the provider cases. Note that these metrics can be normalized in different ways. For example, the count-oriented metrics may be normalized by the size of the provider catalog, and / or by the number of users, etc. For simplicity, we omit a complete enumeration of all such variants here. Note also that a provider might be interested in their rank or scoring relative to other providers. For example, an Exposure value of 600 might be more meaningful if the provider is also told that this value ranks 3rd across all providers or that it is 1.2 standard deviations above the mean value.

\begin{table}[tbh]
	\centering
\begin{tabular}{lp{2.0in}p{1.25in}}
Type & Formula & Explanation \\ \hline
Exposure(p) & $\sum_{L_i \in \mathcal{L}}{\sum_{j \in L_i}{\mathbbm{1}(j \in I_p)}}$ & Count the number of recommendations given across all of $p$'s items. \\ \hline
Hits(p) & $\sum_{L_i \in \mathcal{L}}{\sum_{j \in L_i}{\mathbbm{1}(j \in I_p \wedge r_{ij} \in T)}}$ & Count the number of hits in recommendation lists for all of $p$'s items. \\ \hline
Reach(p) & $\sum_{L_i \in \mathcal{L}}{\mathbbm{1}(|I_p \cap L_i| > 0 )}$ & Count how many users get at least one $i_p$ item recommended. \\ \hline
TargetReach(p) & $\sum_{L_i \in \mathcal{L}}{\mathbbm{1}(|I_p \cap L_i| > 0 \wedge g_p(i) )}$ & Count how many users in $p$'s target set get at least one $i_p$ item recommended. \\ \hline
PAccuracy(p,m) & $[\sum_{r_{ij} \in T_p}{m(r_{ij}, \hat{r}_ij)}]/{|T_p|}$ & Average metric $m$ score for predictions of $p$'s items. \\
\hline

\end{tabular}
\caption{Examples of provider metrics.  Let $p$ be a given provider, and $i_p \in I_p$ an item associated with $p$. Let $\mathcal{L} = {L_0, L_1, ..., L_n }$ be the recommendation lists calculated for $n$ users. Let $T$ be set of $r_{ij}$ ratings in the test set over which $\mathcal{L}$ is calculated. Let $T_p$ be provider $p$'s subset of $T$: $T_p = \{r_{ij}: r_{ij} \in T \wedge i \in I_p\}$. Let $\mathbbm{1}$ be the indicator function. Let $m(r_{ij},\hat{r}_{ij})$ be an accuracy-oriented evaluation metric (such as RMSE) that evaluates a predicted rating $\hat{r}_{ij}$ relative to a known value $r_{ij} \in T$. Let $g_p(i)$ be a boolean function that returns true if user $i$ is in the target market of provider $p$. }\label{tab:provider}
\end{table}

\subsubsection{System metrics}
As noted above, the system utility may in many cases be a simple aggregation of the utilities of other parties, the $S_a$ case. For example, in a simple commission-oriented arrangement, the profit to the system might be some weighted aggregate of the \textit{Hits} metric, taking item price and commission rate into account. However, other cases arise where the system has its own targeted utility framework that is not a simple aggregate of those of users or providers. This is the $S_t$ case.

An important $S_t$ context is algorithmic fairness discussed in Section~\ref{sec:far}. In general, we should not expect that providers will care if the system is fair to others as long as it provides them with good outcomes. Any fairness considerations and related metrics will therefore be ones defined by system considerations. For example, we can define our provider metrics to map to providers with and without sensitive features, and $P(Y=1)$ as Reach(p)$/|\mathcal{L}|$. Then, a proportional impact metric such as $\Pr[\hat{Y}{=}1 | S{=}0] / \Pr[\hat{Y}{=}1 | S{=}1] \rightarrow 1$ can be defined to see how closely a particular set of recommendation results tracks the desired fairness outcome where this ratio approaches 1.

Other discussions of system utilities are relatively sparse in the multistakeholder recommendation literature. As noted in Section~\ref{sec:review-profit} above, considerations such as customer lifetime value are candidates for $S_t$ metrics, as they are not simply reducible to the utilities of other stakeholders. Some of the applications discussed below, such as educational recommendation, also present some interesting challenging for defining and applying system metrics.

\section{Research Directions}
    \label{sec:directions}
While there are a number of examples of notable research results in multistakeholder recommendation, a number of important unsolved challenges remain. In this section, we examine some important research directions in which we expect future progress.

\subsection{Algorithms}
As noted above, existing work has explored some algorithmic approaches to multistakeholder recommendation. Two approaches can be identified: the first situates the multistakeholder problem within the core recommendation generation function as a type of multi-objective optimization, the second applies multistakeholder considerations after an initial set of recommendations has been generated.

The multi-objective approach, typified by the work discussed here in Section~\ref{sec:far:process:algorithm}, builds a loss function that incorporates multiple objectives and attempts to learn a recommendation function that is sensitive both to a standard accuracy-oriented objective, which can be understood as a $C_p$ consideration, and to an objective that is oriented towards some other stakeholder. In Section~\ref{sec:far:process:algorithm}, this is a fairness objective belonging to the system. Important research challenges remain in formulating and applying multiple objectives across the wide range of multistakeholder applications.

The second algorithm type, as demonstrated in \ref{sec:implementing}, is one that employs an existing recommendation algorithm to generate recommendations (again, this is understood as the user-oriented aspect of the system) and then other stakeholders' considerations are integrated through a re-ranking process. Such systems have the benefit of being modular, so that improvements can be made and analyzed for each part of the algorithm separately. Researchers have built on existing work in information retrieval such as MMR \citep{karako2018image} and xQuad \citep{liu2018personalizing} as well as constraint satisfaction \citep{surer2018multistakeholder} and probabilistic soft logic \citep{farnadi2018fairness}. We expect multistakeholder-oriented re-ranking will remain an important research direction for the field.

One discernible trend in algorithmic research in recommender systems has been the move from narrower objectives for recommendation algorithms to broader ones. Initially, point-wise accuracy metrics were developed, which evolved into pair-wise and list-wise metrics, and more recently to considering interactions extended in time. Multistakeholder recommendation raises the possibility of broadening the objective yet again towards optimizing over the entire set of recommendations delivered. The approach in \citep{surer2018multistakeholder} is one step in this direction, as it formulates the re-ranking problem as approximating constraint satisfaction over all of the recommendation lists generated for the test data. While the interactive requirements of recommendation may seem to argue against the computation of a global optimum, many applications generate and cache recommendations and would be a good match for this kind of algorithm. 

\subsection{Applications}
The pattern of multistakeholder recommendation can be observed in variety of different applications. The prior discussion has highlighted existing research in the reciprocal domains of job recommendation and online dating, in the value-aware environments in e-commerce and multisided platforms, and in the environments where fairness considerations apply. There are many additional areas of application for multistakeholder recommendation, some of which are listed here.  

\begin{description}
    \item [Education:]
    Depending on the environment, recommendation of educational content may have a multistakeholder aspect. For example, there may be tension between the interests of students who want to pursue familiar content and those of the educational system that may be interested in producing students with a well-rounded range of experiences. When multiple educational providers are involved, there may be provider considerations as well~\citep{burke_educational_2016}. Sector: $\langle C^*_p, P^-_*, S_t \rangle$
    
    \item [Philanthropy:]
    Commerce-oriented multisided platforms are obvious examples where multistakeholder considerations are important. However, there are also multisided platforms that have philanthropic aims. The crowd-sourced microlending platform Kiva.org is such an example where fairness-aware recommendation has been applied~\citep{burke2018balanced}. Sector: $\langle C^*_p, P^-_n, S_t \rangle$
     
    \item [Tourism:]
    Another example of recommender systems involving multiple stakeholders is tourism. For example when a travel recommender system recommends a destination or a travel package to a user, the stakeholders that are involved include the traveler, the airlines (or any other transportation provider), the host (destination) and also the system. The hotel recommendation system in Section~\ref{sec:implementing} shows some of the multisided nature of interactions in travel. Peer-to-peer travel services like AirBnB may also have reciprocal aspects. Sector: Sector: $\langle C^*_p, P^*_*, S_a \rangle$

    \item [News recommendation:]
    News recommendation can be viewed as strictly a matter of personalizing for user taste, but there are considerations -- such as public service goals or regulatory requirements -- that might require fairness, balance or other system objectives. Sector: $\langle C^*_p, P^-_n, S_t \rangle$
    
    \item [Social media:]
    In social media platforms, users get a variety of different content merged as a ranked content stream, that can be understood as a set of recommendations. For example, on Facebook, users get friends' posts as one type of recommendation and ads as another type. Thus, we have multiple types of providers and a task of balancing the content of the feed so that multiple objectives are met. Sector: $\langle C^*_p, P^-_*, S_{a,t} \rangle$
    
\end{description}

As this set makes clear, the scope of multistakeholder recommendation is quite broad and incorporates systems of societal importance. It may be inevitable that, as recommender systems move further into applications with more significant stakes for individuals and society, it becomes more and more necessary that they serve a multiplicity of purposes, something that cannot be achieved with a strict focus on the end user.

\subsection{Explanation / Transparency}
Exploring how recommendation explanation could be done in a multistakeholder environment is also another direction for future research. Explanation is an important factor in recommendation interfaces, helping users understand how a given recommendation relates to their interests. It has been shown that explanations can enhance users' likelihood of adopting a given recommendation.

Multistakeholder recommendation poses some interesting challenges for recommendation explanation. First, there is the issue of complexity: a recommendation produced by a multistakeholder system will, by necessity, be one that incorporates multiple factors in its production, and therefore any explanation will be more complex than what would be needed if user preferences were the only consideration. In addition, there is the fact that the objectives of some of the other stakeholders may be in conflict with those of the user. In some contexts, one could imagine users finding it objectionable that their preferences are being downplayed in favor of others' interests. E-commerce sites that confront this problem often label items in recommendation lists as ``promoted'' or ``sponsored'' when they are being displayed because of provider consideration. It is more difficult to do this when a global optimization algorithm is being applied, as all results will potentially be influenced by the full set of stakeholders. Producing acceptable explanations in such contexts is an interesting challenge, but a good solution may be necessary to make multistakeholder recommendations broadly useful.

\section{Conclusion}
    Multistakeholder recommendation is an important development in the evolution of the recommender systems field, as researchers widen their view of those impacted by the results recommender systems produce. This is a natural progression from the initial academic research prototypes to today's fielded systems, key elements of online applications, with millions of users. It is not surprising that systems occupying key positions in complex commercial and social environments should have to answer to many masters.

While multistakeholder issues have surfaced regularly in the history of recommender systems research, the recognition of common threads and research questions has been a more recent occurrence. This article has presented a synthesis of the landscape of this research past and present, demonstrated some important current applications, and raised important questions for future investigation.

\bibliographystyle{plainnat}


\end{document}